\documentclass[11pt]{article}\pdfoutput=1
\usepackage{cite}
\usepackage{amsmath,amsfonts,amssymb}
\usepackage[small,bf,hang]{caption}
\usepackage{slashed}
\usepackage{amsmath}
\usepackage{latexsym,epsfig}
\usepackage{arydshln}

\usepackage[vcentermath]{youngtab}

\def\hybrid{
        \topmargin -20pt
        \oddsidemargin 0pt
        \headheight 0pt \headsep 0pt
        \textwidth 6.25in 
        \textheight 9.5in 
        \marginparwidth .875in
        \parskip 5pt plus 1pt \jot = 1.5ex}

\hybrid

 \csname
@addtoreset\endcsname{equation}{section}

\def\moth{\mathsurround=0pt}
\newdimen\zo \zo=0pt

\def\tick{\leaders\hrule height 0.5ex depth 0pt \hskip 0.5pt}
\def\upboxfill{$\moth \setbox\zo\hbox{\tick}%
  \hskip 3pt\hbox to 0pt{$\tick$\hss}\hrulefill \hbox to 7.5pt{$\tick$\hss}$}

\def\dtick{\leaders\hrule height .34pt depth 0.5ex \hskip 0.5pt}
\def\downboxfill{$\moth \setbox\zo\hbox{\dtick}%
  \hskip 2pt\hbox to 0pt{$\dtick$\hss}\hrulefill \hbox to 2pt{$\dtick$\hss}$}

\def\bec{\begin{center}}
\def\ec{\end{center}}

 \def\det{{\rm det\,}}
\def\be{\begin{equation}}
\def\ee{\end{equation}}
\def\bea{\begin{eqnarray}}
\def\eea{\end{eqnarray}}
\def\ba{\begin{array}}
\def\ea{\end{array}}

\begin{document}

\begin{titlepage}
\rightline{}
\rightline{October 2018}
\rightline{HU-EP-18/31}
\begin{center}
\vskip 1.5cm
 {\Large \bf{Reviving  3D  ${\cal N}=8$  
superconformal field theories }}
\vskip 1.7cm

{\large\bf {Olaf Hohm${}^{1,2}$ and Henning Samtleben${}^3$}}
\vskip 1.6cm

{\it ${}^1$ Institute for Physics, Humboldt University Berlin,\\
 Zum Gro\ss en Windkanal 6, D-12489 Berlin, Germany}\\
\vskip .1cm

{\it ${}^2$
Simons Center for Geometry and Physics, Stony Brook University,\\
Stony Brook, NY 11794-3636, USA}\\
ohohm@physik.hu-berlin.de, ohohm@scgp.stonybrook.edu
\vskip .2cm

{\it ${}^3$ Univ Lyon, Ens de Lyon, Univ Claude Bernard, CNRS,\\
Laboratoire de Physique, F-69342 Lyon, France} \\
{henning.samtleben@ens-lyon.fr}

\end{center}

\bigskip\bigskip
\begin{center} 
\textbf{Abstract}

\end{center} 
\begin{quote}

We present a Lagrangian formulation for ${\cal N}=8$ superconformal field theories 
in three spacetime dimensions that is general enough to encompass infinite-dimensional 
gauge algebras that generally go beyond Lie algebras. 
To this end we employ Chern-Simons theories based on Leibniz algebras, which give rise to L$_{\infty}$ algebras
and are defined on the dual space $\frak{g}^*$ of a Lie algebra $\frak{g}$ by means of 
an embedding tensor map $\vartheta :\frak{g}^*\rightarrow \frak{g}$. 
We show that for the Lie algebra $\frak{sdiff}_3$ of volume-preserving diffeomorphisms 
on a 3-manifold there is a natural embedding tensor 
defining a Leibniz algebra on the space of one-forms. 
Specifically, we show that the cotangent bundle to any 3-manifold with a volume-form 
carries the structure of a (generalized) Courant algebroid. 
The resulting ${\cal N}=8$ superconformal field theories are  shown to be equivalent to 
Bandos-Townsend theories. We show that the theory based on $S^3$ is an infinite-dimensional 
generalization of the Bagger-Lambert-Gustavsson model that in turn 
is a consistent truncation of the full theory. 
We also review a Scherk-Schwarz reduction on $S^2\times S^1$, which gives the super-Yang-Mills 
theory with gauge algebra $\frak{sdiff}_2$, and we construct massive deformations.

\end{quote} 
\vfill
\setcounter{footnote}{0}
\end{titlepage}

\tableofcontents


\section{Introduction}

Conformal field theories in three dimensions (3D) with large amounts of supersymmetry are interesting for several reasons. 
These include the description of multiple M2 branes, examples of the AdS$_4$/CFT$_3$ correspondence 
and as (possibly solvable) toy models for conformal fixed points in condensed matter systems.
There is no a priori reason why such superconformal field theories (SCFT) should have a Lagrangian formulation, 
but there are examples of Lagrangian SCFTs in 3D, notably the Aharony-Bergman-Jafferis-Maldacena (ABJM) theories \cite{Aharony:2008ug}, 
which are Chern-Simons-matter theories featuring ${\cal N}=6$ supersymmetry and non-abelian gauge groups such as $SU(N)\times SU(N)$. 
These theories have been employed extensively 
for applications of the type mentioned above, but they generally do not exhibit the largest amount of supersymmetry 
expected to be possible, which is ${\cal N}=8$ corresponding to 16 real supercharges. 
There is one non-trivial 3D ${\cal N}=8$ SCFT Lagrangian with gauge group $SO(4)=SU(2)\times SU(2)$, the Bagger-Lambert-Gustavsson (BLG) model  
\cite{Bagger:2007jr,Gustavsson:2007vu}, but it has been proven that, under physically reasonable assumptions, this model 
cannot be extended to include gauge groups of arbitrary rank, such as $U(N)$ for arbitrary $N$ \cite{Papadopoulos:2008sk,Gauntlett:2008uf}. 
Since such gauge groups are needed for the applications discussed above, where $N$ would be related to the 
number of M2 branes, or where one would like to take the planar or $N\rightarrow \infty$ limit, the BLG model has been of limited use, 
and it remains puzzling why there is such an isolated point in the `space of Lagrangian SCFTs'.  

In this  paper we revisit 3D ${\cal N}=8$ theories with the goal to include infinite-dimensional gauge groups, 
which are not covered by the no-go results of  \cite{Papadopoulos:2008sk,Gauntlett:2008uf}. 
There is in fact a generalization of the BLG model, due to Bandos and 
Townsend \cite{Bandos:2008jv}, 
in which the gauge group is given by the volume-preserving diffeomorphisms of a 3-manifold (subject to certain topological constraints), 
and which takes the structural form of a Chern-Simons-matter theory  in $3+3$ dimensions. 
Surprisingly, since its discovery in 2008 this theory has attracted little attention. Perhaps one reason for this is that 
the formulation of \cite{Bandos:2008jv} is not manifestly local: in order to write  a Chern-Simons-like action 
new fields are needed that can only non-locally be expressed in terms of the fundamental fields.  
Accordingly, the theory has been termed `exotic' in \cite{Bandos:2008jv}, because 
it is not obtained as an abstract Yang-Mills theory based on a Lie algebra. 
As one of the main results of this paper, we present a universal and local formulation of ${\cal N}=8$ superconformal Lagrangians  
that is applicable to infinite-dimensional gauge algebras and that contains all currently known ${\cal N}=8$ SCFTs, including Bandos-Townsend theories, 
as special cases. We will then show that the Bandos-Townsend theory based on $S^3$
provides  a generalization of the BLG model, in which 
the $SO(4)$ gauge group becomes part of an infinite-dimensional gauge symmetry, realized on an infinite number of fields, 
and of which the BLG model is a consistent truncation.  In addition, it was already discussed in \cite{Bandos:2008jv} that 
for the $S^2\times S^1$ theory Scherk-Schwarz reduction on $S^1$ 
yields a 3D ${\cal N}=8$ super-Yang-Mills theory with gauge group SDiff$_2$, which in turn can be viewed as the $N\rightarrow \infty$
limit of $SU(N)$ gauge theories (see, e.g., \cite{Floratos:1988mh}). A web of these theories and how they can be derived from the SDiff$_3$ `parent theory' 
is indicated in the figure below. 
\begin{figure}[hbt]
   \centering
   \includegraphics[width=11.6cm]{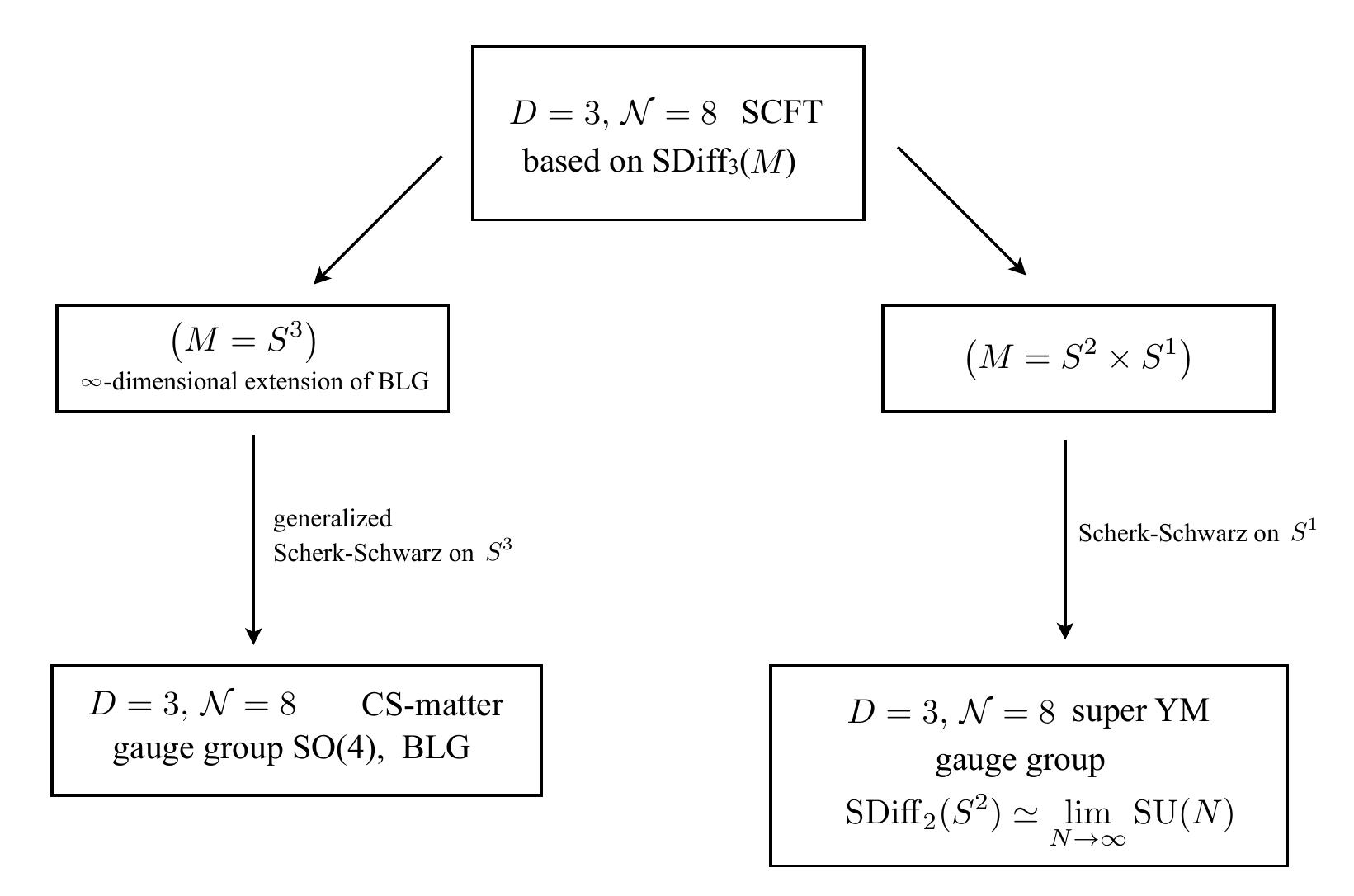}
     \caption{{\small Web of various theories}}
   \label{Figure:E6}
\end{figure}

The formulation to be developed here 
 is based on novel algebraic structures beyond Lie algebras, 
using in a simpler setting 
techniques that have recently been developed in `exceptional field theory' \cite{Hohm:2013pua,Hohm:2013vpa,Hohm:2013uia,Hohm:2014fxa}, 
the U-duality covariant formulation  of maximal supergravity, 
which in turn is based on double field theory \cite{Siegel:1993th,Hull:2009mi,Hull:2009zb,Hohm:2010jy,Hohm:2010pp} 
and generalized geometry \cite{TCourant,LieXu,roytenberg-weinstein,roytenberg}. 
Specifically, we employ the recent construction of Chern-Simons theories in \cite{Hohm:2018ybo,Hohm:2017wtr} 
based on so-called Leibniz-Loday algebras \cite{LODAY}, which are generalizations of Lie algebras that give rise to L$_{\infty}$ 
algebras \cite{Zwiebach:1992ie} (see \cite{Hohm:2017pnh} for an introduction to L$_{\infty}$ 
algebras). 
In the cases of interest the Leibniz-Loday  algebras are derived from a Lie algebra by means of an embedding tensor formulation \cite{Hohm:2018ybo}, 
which in turn generalizes techniques to infinite-dimensional contexts  
that were introduced in a finite-dimensional setting in gauged supergravity  \cite{Nicolai:2000sc,deWit:2002vt,deWit:2003ja}.

For the convenience of the reader, we now summarize the general data needed in order 
to define an ${\cal N}=8$ SCFT and what constraints they are subject to. The general data are
 \begin{itemize}
  \item a (generally infinite-dimensional) Lie algebra $\frak{g}$;  
  \item an embedding tensor map $\vartheta:\, \frak{g}^* \ \rightarrow \ \frak{g}$, where $\frak{g}^*$ is the dual space to $\frak{g}$; 
  \item a $\frak{g}$ 
  representation $\rho_{v}:R\rightarrow R$, $v\in \frak{g}$, on a space $R$ with $\frak{g}$-invariant bilinear form $\langle\cdot\,, \cdot\rangle$. 
 \end{itemize} 
Provided these data satisfy certain constraints to be discussed shortly,\footnote{Here we use a rather liberal notion of `dual space'. Since in the interesting cases $\frak{g}$
will be infinite-dimensional it is generally a subtle question how the dual space $\frak{g}^*$ should be defined precisely. 
All we need is that $\frak{g}^*$ is a representation space of $\frak{g}$ (coadjoint representation), with 
an invariant  pairing $\omega(v)\in \mathbb{R}$ 
between a vector $v\in \frak{g}$ and a covector $\omega\in \frak{g}^*$. 
In particular, the pairing does not even need to be non-degenerate.} 
they define algebraic structures including Leibniz-Loday algebras that are sufficient in order to 
define a Lagrangian for an ${\cal N}=8$ superconformal field theory.
Specifically, the embedding tensor $\vartheta$ defines a bilinear, not necessarily antisymmetric product on $\frak{g}^*$ by 
    \be\label{LeibnizFirst}
    \omega \circ \eta \ \equiv \ {\rm ad}^*_{\vartheta(\omega)}\eta\;, 
   \ee
  where  $\omega, \eta\in \frak{g}^*$, and ${\rm ad}^*$ is the coadjoint representation of $\frak{g}$ on $\frak{g}^*$. 
Alternatively, the embedding tensor can be viewed as a map 
 $\Theta:\frak{g}^*\otimes \frak{g}^*\rightarrow \mathbb{R}$, defined by  
 \be\label{bigThetafirst} 
  \Theta(\omega,\eta) \ \equiv \ -\omega(\vartheta(\eta))\;, 
 \ee 
 in terms of the natural pairing between $\frak{g}^*$ and  $\frak{g}$. We assume this map to be symmetric. 
 The embedding tensor $\vartheta$ in general is not an isomorphism, and therefore the Lie algebra structure on $\frak{g}$ 
 cannot be transported to a Lie algebra structure on $\frak{g}^*$. However, provided the following \textit{quadratic constraint} is satisfied, 
 it is transported to a Leibniz-Loday structure on $\frak{g}^*$ \cite{Strobl}, which is sufficient in order to define a Chern-Simons gauge theory. 
 In addition, in order to define a complete ${\cal N}=8$ Lagrangian, an invariant 4-tensor 
 on the representation space $R$ needs to satisfy a \textit{linear constraint}, and thus in total the above data are 
 subject to the following constraints: 
 \begin{itemize}
  \item[(1)] \textit{Quadratic constraint}:\\
  The product (\ref{LeibnizFirst}) needs to be compatible with the Lie bracket $[\cdot,\cdot]$ on $\frak{g}$ in that 
   \be\label{quadconstr0}
    \vartheta(\omega\circ \eta) \ = \ \big[\vartheta(\omega), \vartheta(\eta)\big]\;. 
   \ee 
 \item[(2)] \textit{Linear constraint:} \\
 With the above data, the representation space $R$ universally carries an invariant 4-tensor, which 
 is defined in terms of (\ref{bigThetafirst}) and the map $\pi:R\wedge R\rightarrow \frak{g}^*$ 
 defined for $\phi_1, \phi_2\in R$, $v\in \frak{g}$ by $\pi(\phi_1, \phi_2)(v)=\langle \phi_1, \rho_{v}\phi_2\rangle$, 
 which is antisymmetric due to the invariance of the bilinear form. 
 The invariant 4-tensor given by 
   \be\label{4tensorintro}
    \{\phi_1,\phi_2, \phi_3, \phi_4\} \ \equiv \ \Theta(\pi(\phi_1,\phi_2), \pi(\phi_3,\phi_4))\;, 
  \ee
needs to be totally antisymmetric.  
  
 \end{itemize}

In the case that the Lie algebra $\frak{g}$ is finite-dimensional the above provides an invariant (or index-free) formulation of  the conventional 
embedding tensor formalism \cite{Nicolai:2000sc,deWit:2002vt,deWit:2003ja}, 
which has already been used to construct  conformal and non-conformal supersymmetric  3D theories and to derive 
them from the corresponding gauged supergravities \cite{Bergshoeff:2008cz,Bergshoeff:2008ix,Bergshoeff:2008bh}.  
It should be pointed out that the original formulation 
by Bagger and Lambert was based on a so-called `3-algebra', a trilinear bracket satisfying a quadratic constraint (`the fundamental identity') 
\cite{Bagger:2007jr}. This is contained in the above formulation since, given the invariant metric on $R$, one can 
view the 4-tensor (\ref{4tensorintro}) as a map (3-bracket) $R\wedge R\wedge R\rightarrow R$, 
obeying  the fundamental identity as a consequence of the quadratic constraint (\ref{quadconstr0}). 
Moreover, picking the Lie algebra $\frak{so}(n)$, which is a global symmetry of the free ${\cal N}=8$ SCFTs with $n$ matter multiplets, 
and taking $R$ to be 
the vector representation, the `3-algebra' formulation is fully equivalent to the embedding tensor formulation. 
Indeed, in this case one has the isomorphism $\frak{so}(n)\simeq R\wedge R$, which is given 
by the above map $\pi$, and therefore the 3-bracket can be derived from the embedding tensor and vice versa. 
In contrast, for more general representations, and certainly for infinite-dimensional Lie algebras, the map $\pi$ will generally not be invertible.  
One may always construct a `3-bracket' from the 4-tensor  (\ref{4tensorintro}), but conversely the embedding tensor, 
which is needed in particular to write the Chern-Simons action, generally cannot be 
reconstructed from the 3-bracket. Thus, the embedding tensor formulation appears to be more general.

In the remainder of this introduction we will briefly outline how the SCFTs based on the algebra $\frak{sdiff}_3$ of 
volume-preserving diffeomorphisms of a 3-manifold are described within the above framework, 
and  in which sense such theories can be viewed as `higher' gauge theories. 
We first note that one cannot define directly a Yang-Mills or Chern-Simons theory for a Lie algebra of (infinitesimal) diffeomorphisms, 
because this requires an invariant bilinear form, which for diffeomorphism algebras generally do not exist. Consequently, $\frak{g}$
and $\frak{g}^*$ are typically not equivalent. 
For the special case of volume-preserving diffeomorphisms of a 3-manifold (subject to certain topological assumptions) 
there actually is an invariant bilinear form, but 
it is non-local. A manifestly local formulation can instead be obtained by 
defining the gauge structure on $\frak{g}^*$, which we here take to be the space of one-forms on the 3-manifold. 
There is then a natural embedding tensor, for which the symmetric bilinear form (\ref{bigThetafirst}) is given 
by $\Theta(\omega,\eta)\propto \int \omega\wedge {\rm d}\eta$, which is manifestly diffeomorphism invariant.

In order to further elucidate  the emergence of `higher' algebra structures 
let us point out that our choice of $\frak{g}^*$ is larger than 
is conventional for the `dual space'. The dual space is more conventionally  defined 
as the space of one-forms modulo exact one-forms, so that the pairing between vectors and covectors 
is non-degenerate. Then, however, one would have to work with fields taking values in equivalence classes. 
If one is willing to do so, there is a  conventional Lie algebra formulation, but one has to abandon manifest locality, 
as we will discuss in more detail. The route taken here, which is common in physics, is to work with fields living on the larger space but 
to impose a gauge redundancy. Then there is a perfectly local formulation, but with a gauge structure that is governed by an L$_{\infty}$ algebra rather than  a 
Lie algebra. For the case at hand, we will show that the cotangent bundle $T^*M$ for any 3-manifold $M$  
equipped only with a volume form provides (a mild generalization of) a Courant algebroid, which in turn gives rise to an 
L$_{\infty}$ structure \cite{roytenberg-weinstein}.  
Thus, the gauge algebras of the SCFTs based on $\frak{sdiff}_3$ are governed in particular by
a non-trivial `3-bracket' encoding the failure of the `2-bracket', the antisymmetric part of the Leibniz product (\ref{LeibnizFirst}), 
to satisfy the Jacobi identity. Taking $R$ to be the 
space of functions on the 3-manifold,  there is also the 4-tensor (\ref{4tensorintro}) and hence a trilinear bracket on $R$, 
which turns out to be the Nambu bracket \cite{Nambu:1973qe}, but it should be emphasized that the latter bracket is \textit{not} directly related to the 
3-bracket defining the gauge algebra; in fact, they are not even defined on the same space.
To illuminate this point, consider the BLG model, for which the trilinear bracket, the `3-algebra' of \cite{Bagger:2007jr}, 
is given by the invariant epsilon tensor of $SO(4)$, i.e., $\{\phi_1,\phi_2,\phi_3\}^a=\varepsilon^{abcd}\phi_{1b}\phi_{2c}\phi_{3d}$. 
The gauge algebra, on the other hand, is the Lie algebra structure on $\frak{so}(4)^*$ that is transported via the embedding tensor 
$\vartheta(\tilde{t}_{ab})  =  \tfrac{1}{2}\varepsilon_{abcd} t^{cd}$ from 
the Lie algebra $\frak{so}(4)$, where $t^{ab}$ and $\tilde{t}_{ab}$ are the generators and dual generators, respectively.  
Being a Lie algebra, this trivially defines an L$_{\infty}$ algebra whose 3-bracket is identically zero.

The rest of this paper is organized as follows. In sec.~2 we develop the general formulation of ${\cal N}=8$ SCFTs, 
starting from a Lie algebra $\frak{g}$ and constructing, via an embedding tensor, a Leibniz-Loday algebra that is sufficient 
in order to define gauge theories. In particular, we show that these general structures are sufficient in order to prove 
${\cal N}=8$ supersymmetry. In sec.~3 we specialize to the Lie algebra of volume-preserving diffeomorphisms on a 3-manifold,
and we show that there is a natural embedding tensor satisfying all constraints. We then show that the resulting theories 
are equivalent, modulo some topological assumptions, to the Bandos-Townsend theories. 
In sec.~4 we consider truncations and deformations of the general framework. Specifically, we consider the mode expansion of the $\frak{sdiff}_3$
SCFTs  for $S^3$ and show that the BLG model is a consistent (generalized Scherk-Schwarz) truncation. 
Conversely, the full Bandos-Townsend theory for $S^3$ provides an infinite-dimensional generalization of the BLG model. 
We also review  the $S^2\times S^1$ model, whose Scherk-Schwarz reduction on $S^1$ 
yields a 3D ${\cal N}=8$ super-Yang-Mills theory with gauge group SDiff$_2$, and we discuss massive deformations that preserve ${\cal N}=8$ supersymmetry. 
In the conclusion section we discuss possible generalizations.

\section{Maximal 3D superconformal field theories } 

We define the 3D ${\cal N}=8$ superconformal field theory for the general data given in the introduction. 
In the first subsection we review how an embedding tensor satisfying the quadratic constraint defines a Leibniz-Loday 
algebra and a gauge invariant Chern-Simons action. In the second subsection we introduce a representation $R$ 
with invariant metric and show that with these data one can universally define an invariant 4-tensor, which in turn 
satisfies a quadratic identity as a consequence of the quadratic constraint. With these structures we are able to 
define in the third subsection the complete ${\cal N}=8$ Lagrangian and prove supersymmetry.

\subsection{Embedding tensor and Leibniz-Chern-Simons theory }

Our starting point is a (generally infinite-dimensional) Lie algebra $\frak{g}$, with Lie brackets $[\cdot,\cdot]$, 
and its dual space $\frak{g}^*$.  As mentioned in the introduction, we only need a weak notion of dual space, 
meaning that it is a representation space (coadjoint representation) with a pairing between covector and vector 
(giving a real number) that is invariant under the combined adjoint and  
coadjoint action. Writing $v, w\in \frak{g}$ for vectors and $\omega, \eta  \in \frak{g}^*$ 
for covectors, the pairing is written as $\omega(v)\in\mathbb{R}$. 
The coadjoint representation is denoted by ${\rm ad}_{w}^*$, and denoting collectively  by $\delta_w$
the adjoint and coadjoint action we have 
 \be\label{pairinfinv}
  \delta_{w}(\omega(v)) \ \equiv \ ({\rm ad}^*_{w}\omega)(v) + \omega([w, v])
 \ = \ 0\;. 
 \ee
Moreover, being a representation, the coadjoint action  also satisfies $[{\rm ad}^*_{v}, {\rm ad}^*_{w}]={\rm ad}_{[v,w]}^*$. 

We now assume the existence of an \textit{embedding tensor}, a linear map 
\be
 \vartheta\,: \quad \frak{g}^*\quad \rightarrow\quad \frak{g}\;, 
\ee
that in general is not invertible. Generally, since $\vartheta$ is not an isomorphism, the Lie algebra structure on $\frak{g}$ 
cannot be transported to a Lie algebra structure on $\frak{g}^*$. There is, however, a natural bilinear operation 
defined in (\ref{LeibnizFirst}), 
  \be\label{LeibnizMAIN}
    \omega \circ \eta \ \equiv \ {\rm ad}^*_{\vartheta(\omega)}\eta\;. 
   \ee
We will now show that this product defines a Leibniz-Loday algebra, which means that the 
following Jacobi-type identity holds 
\be\label{JacobiLeibniz}
 \begin{split}
  (\omega_1\circ \omega_2)\circ \omega_3 +  \omega_2\circ (\omega_1\circ \omega_3)   - \omega_1 \circ (\omega_2\circ \omega_3) 
   \ = \ 0 \;, 
 \end{split} 
 \ee
provided the embedding tensor 
satisfies the quadratic constraint (\ref{quadconstr0}), 
   \be\label{quadconstr}
    \vartheta(\omega\circ \eta) \ = \ \big[\vartheta(\omega), \vartheta(\eta)\big]\;. 
   \ee 
The proof proceeds by noting that the left-hand side of the Leibniz relation (\ref{JacobiLeibniz}) can 
be rewritten as 
  \be
 \begin{split}
  {\rm ad}^*_{\vartheta(\omega_1\circ \omega_2)}\omega_3 -\big [ {\rm ad}^*_{\vartheta(\omega_1)}, {\rm ad}^*_{\vartheta(\omega_2)}\big]\omega_3 
  \ =  \ {\rm ad}^*_{\vartheta(\omega_1\circ \omega_2)-\big[\vartheta(\omega_1), \vartheta(\omega_2)\big]}\omega_3 \ = \ 0 \;, 
 \end{split} 
 \ee
using that ${\rm ad}^*$ is a representation of $\frak{g}$.  

The Leibniz relation (\ref{JacobiLeibniz}) can be interpreted as saying that the action of $\frak{g}^*$ on itself, defined by 
 \be\label{Leibnizaction}
  \delta_{\omega}\eta \ \equiv \ \omega\circ \eta\;, 
 \ee   
closes in that $[\delta_{\omega_1}, \delta_{\omega_2}]=\delta_{\omega_1\circ \omega_2}$. 
Moreover, defining the symmetric and antisymmetric  parts of the Leibniz product,    
    \be\label{symmantisymm}
  \begin{split}
   [\omega,\eta] \ \equiv \ \tfrac{1}{2}(\omega\circ \eta-\eta\circ \omega)\;, \\
   \{\omega,\eta\} \ \equiv \ \tfrac{1}{2}(\omega\circ \eta+\eta\circ \omega)\;, 
  \end{split}
 \ee 
this closure relation can also be written as $[\delta_{\omega_1}, \delta_{\omega_2}]=\delta_{[\omega_1, \omega_2]}$ 
as a consequence of the left-hand side being manifestly antisymmetric. 
As a consistency check we note that symmetrizing (\ref{quadconstr}) in $\omega, \eta$, the right-hand side  vanishes, 
and we thus learn  
 \be\label{thetasymtriv}
  \vartheta(\{\omega, \eta\}) \ = \ 0\;. 
 \ee
Thus, the symmetrized Leibniz product lives in the kernel of $\vartheta$ and hence 
defines a trivial action. As in \cite{Hohm:2018ybo} we find it convenient 
to represent the ideal of such trivial vectors as the image of a linear operator ${\cal D}$, so that we can write 
\be\label{DREL}
  \{\omega,\eta\} \ = \  \tfrac{1}{2}{\cal D}(\omega\bullet \eta)\;, 
 \ee
with a new bilinear operation $\bullet$. The Leibniz action (\ref{Leibnizaction}) is then trivial for ${\cal D}$-exact gauge parameters, 
$\delta_{{\cal D}\chi}=0$. 

The relation between the Lie algebra on $\frak{g}$ and the Leibniz algebra on $\frak{g}^*$ can be summarized   as follows: 
If the symmetric part of the Leibniz product is non-zero the embedding tensor map $\vartheta$ has a non-trivial kernel, 
in particular is not an isomorphism, in agreement with the Leibniz algebra not being a Lie algebra. 
Conversely, if the symmetric part of the Leibniz product vanishes, the relation (\ref{JacobiLeibniz}) 
reduces to the Jacobi identity, and the Leibniz algebra becomes a Lie algebra. The quadratic constraint 
(\ref{quadconstr}) then says that $\vartheta$ 
is a Lie algebra homomorphism.

For the subsequent constructions it will be important to also interpret the embedding tensor as a symmetric tensor 
 \be
  \Theta:\;\;\frak{g}^*\otimes \frak{g}^* \ \rightarrow \ \mathbb{R}\;, 
 \ee  
defined by  
 \be\label{DefTheta} 
  \Theta(\omega,\eta) \ \equiv \ -\omega(\vartheta(\eta))\;, 
 \ee 
using the pairing between vector and covector on the right-hand side. 
This tensor is invariant under the Leibniz action (\ref{Leibnizaction}) as a consequence of the quadratic constraint 
(\ref{quadconstr}): 
 \be\label{THETaisINv}
  \begin{split}
   \delta_{\omega}\Theta(\eta_1, \eta_2) \ &\equiv \ \Theta(\omega\circ \eta_1, \eta_2) + \Theta(\eta_1, \omega\circ \eta_2) \\
   \ &= \ -(\omega\circ \eta_1)(\vartheta(\eta_2)) - \eta_1(\vartheta(\omega\circ \eta_2))\\
   \ &= \ -({\rm ad}_{\vartheta(\omega)}^*\eta_1)(\vartheta(\eta_2)) - \eta_1([\vartheta(\omega), \vartheta(\eta_2)])\\
   \ &= \ 0\;, 
  \end{split}
 \ee    
using (\ref{pairinfinv}) in the last step. Moreover, from the symmetry of $\Theta$, the definition (\ref{DefTheta}) 
and (\ref{thetasymtriv}) it immediately follows that 
 \be\label{Thetahigherinv}
  \Theta(\omega, \{\eta_1, \eta_2\}) \ = \ 0\;, 
 \ee
or, equivalently, $\Theta(\omega, {\cal D}\chi)=0$. Thus, $\Theta$ is generally degenerate. 

We are now ready to define a 3D Chern-Simons theory for one-form gauge fields $A_{\mu}\in \frak{g}^*$, taking values 
in the space carrying a Leibniz algebra structure. The gauge transformations are given by 
\be\label{FullDeltaA}
 \delta_{\lambda}A_{\mu} \ = \ D_{\mu}\lambda \  -  \   
 {\cal D}\lambda_{\mu}\;, 
\ee
with covariant derivative $D_{\mu}=\partial_{\mu}-A_{\mu}\,\circ $, where $\lambda\in\frak{g}^*$ and $\lambda_{\mu}$ 
lives in the space in which the bilinear operation $\bullet$ takes values. 
With respect to the above structures, we can write the Chern-Simons action 
 \be\label{CSSS}
  S_{\rm CS} \ =  \ \int {\rm d}^3x\, \varepsilon^{\mu\nu\rho}\,\Theta\big(A_{\mu}, \partial_{\nu}A_{\rho}
  \ -\ \tfrac{1}{3}  A_{\nu}\circ A_{\rho}\big)\;. 
 \ee
 Note that the Leibniz product in the cubic term can be replaced by the antisymmetric bracket, 
 but it should be recalled that this bracket does not define a Lie algebra. However, despite this bracket 
 not satisfying the Jacobi identity, the above action is gauge invariant due to $\Theta$ satisfying (\ref{Thetahigherinv}); see 
 \cite{Hohm:2018ybo} for a more detailed derivation. The general variation of (\ref{CSSS}) w.r.t.~$\delta A_{\mu}$ reads 
   \be\label{generalCSvar}
   \delta_{A}S_{\rm CS} \ = \ \int {\rm d}^3x\,\varepsilon^{\mu\nu\rho}\,\Theta(\delta A_{\mu}, F_{\nu\rho})\;, 
  \ee
 with the field strength, taking values in $\frak{g}^*$,  
  \be\label{fieldSTRENGTH}
   F_{\mu\nu} \ = \ \partial_{\mu}A_{\nu}- \partial_{\nu}A_{\mu} - \big[A_{\mu},A_{\nu}\big]+\cdots, 
  \ee 
 where the ellipsis denotes terms that are ${\cal D}$ exact and hence drop out of (\ref{generalCSvar}) because 
 of (\ref{Thetahigherinv}). (We could introduce two-forms to render (\ref{fieldSTRENGTH}) fully gauge invariant, 
 building a  tensor hierarchy, but this is not needed for the following construction.)

\subsection{Matter fields and invariant four-tensor}

We now introduce matter fields that live in some representation $R$ of the Lie algebra $\frak{g}$. 
We denote the elements of $R$ by $\phi_1, \phi_2$, etc., and the representation by $\delta_{v}\phi \equiv  \rho_{v}\phi$, so that 
 \be\label{rhorep}
  [\rho_{v},\rho_{w}] \ = \ \rho_{[v,w]}\;. 
 \ee
Moreover, we assume that $R$ carries an invariant metric $\langle\,,\rangle$, so that 
$\langle\rho_v \phi_1,\phi_2\rangle=-\langle \phi_1, \rho_v\phi_2\rangle$. 
This metric encodes the kinetic term of the matter fields, and therefore it needs to be positive definite in order to define a ghost-free theory, 
but mathematically the following construction goes through for any signature. 
In the full interacting theory to be constructed the gauge structure is governed by the Leibniz algebra on $\frak{g}^*$ rather than 
the Lie algebra on $\frak{g}$, but the representation $\rho$ induces, by means of the 
embedding tensor, a natural action of the Leibniz algebra on $R$: For $\lambda\in \frak{g}^*$ we set  
 \be\label{LeibnizACTION}
  \delta_{\lambda}\phi \ = \ \rho_{\vartheta(\lambda)}\phi\;. 
 \ee
As a consequence of the quadratic constraint (\ref{quadconstr}) and $\rho$ being a representation, thus satisfying (\ref{rhorep}), 
it immediately follows that these transformations close according to the Leibniz algebra structure, 
$[\delta_{\lambda_1},\delta_{\lambda_2}]=\delta_{\lambda_1\circ \lambda_2}$. Again, we could replace the Leibniz product on 
the right-hand side by the antisymmetric bracket, using that by (\ref{thetasymtriv}) the symmetric part acts trivially.

In the remainder of this subsection we will show that the representation $R$ universally carries a  4-tensor that is invariant 
under the Leibniz action (\ref{LeibnizACTION}). We first note that there is a canonical map 
$\pi :R\otimes R\rightarrow \frak{g}^*$ defined by its pairing with $v\in \frak{g}$,
 \be\label{BestPI}
  (\pi(\phi_1,\phi_2))(v) \ = \ \langle \phi_1, \rho_{v}\phi_2\rangle \ = \ -\langle \rho_{v}\phi_1,\phi_2\rangle\;, 
 \ee
where $\phi_1, \phi_2\in R$, and we used the invariance of the metric $\langle\, , \rangle$ on $R$.  
This map is thus antisymmetric and can hence also be viewed as a map $\pi:R\wedge R\rightarrow \frak{g}^*$. 
As a consequence of this map being defined `intrinsically', in terms of invariant objects, it follows that 
it transforms `covariantly' by the coadjoint representation,\footnote{There is actually a subtlety here 
due to the vector/covector pairing in general being degenerate.  The relations (\ref{BestPI}) may thus 
not uniquely define the map $\pi$. We assume here that $\pi$ has been chosen so that it transforms covariantly 
as in (\ref{PIisCOV}). More generally, any choice of $\pi$ satisfying (\ref{BestPI}) transforms as (\ref{PIisCOV}) 
up to contributions that vanish upon paring with a vector. This ambiguity is, however,  irrelevant for the subsequent construction, since 
$\pi$ will only appear as an argument of the embedding tensor $\Theta$, which is defined in (\ref{DefTheta}) 
by means of the same pairing, so that this ambiguity drops out. 
 }
 \be\label{PIisCOV}
  \delta_{w}\pi(\phi_1,\phi_2) \ \equiv \ \pi(\rho_w\phi_1, \phi_2)+\pi(\phi_1, \rho_w\phi_2)
  \ = \ {\rm ad}_{w}^*(\pi(\phi_1,\phi_2))\;. 
 \ee

We now define a 4-tensor $\{\cdot, \cdot, \cdot, \cdot\}:R^{\otimes 4}\rightarrow \mathbb{R}$ in terms of the embedding tensor $\Theta$ as 
 \be\label{4TENSORR}
  \{\phi_1,\phi_2,\phi_3, \phi_4\} \ \equiv \ \Theta(\pi(\phi_1,\phi_2), \pi(\phi_3,\phi_4))\;. 
 \ee
It follows from the invariance (\ref{THETaisINv}) of $\Theta$ under the Leibniz action and the covariance (\ref{PIisCOV}) 
of the map $\pi$ that the above 4-tensor is invariant under the Leibniz algebra action (\ref{LeibnizACTION}). 
Moreover, this tensor is manifestly antisymmetric in its first two and its second two arguments and symmetric under exchange of the two 
pairs of arguments. Thus, in terms of Young tableaux, it  lives in the symmetric tensor product
 \bea\label{irreducible}
  \left(\hspace{0.2em}{\small \yng(1,1)}\otimes {\small \yng(1,1)}\hspace{0.2em}\right)_{\rm sym}
  \; = \; \;  {\small \yng(1,1,1,1)} \ \oplus \ 
  {\small \yng(2,2)} \; ~. 
 \eea
While supersymmetry requires that this tensor only lives in the totally antisymmetric part, 
the quadratic identities to be derived now hold in all generality. 

Given the invariant metric $\langle\,,\rangle$ on $R$ we can also view this tensor as a  map 
$\{\cdot, \cdot, \cdot\}: R^{\otimes 3}\rightarrow R$ defined by 
 \be\label{3BRACKET}
  \langle \{\phi_1,\phi_2,\phi_3\}, \phi_4 \rangle \ = \ \{\phi_1,\phi_2,\phi_3,\phi_4\}\;. 
 \ee
It should not cause any confusion that we denote both maps by curly parenthesis, since it can always be 
inferred from the number of arguments which map is meant.

We will now prove the `fundamental identity' 
 \be\label{FUNDAMENTALIDD}
  0 \ = \ \sum_{{\rm sym}} \{\{\varphi_1,\varphi_2,\phi_1\}, \phi_2, \phi_3, \phi_4\}\;, 
 \ee 
where the sum implements the symmetries of $\phi_1,\ldots,\phi_4$ according to (\ref{irreducible}), i.e., 
it antisymmetrizes the pairs $[\phi_1,\phi_2]$ and $[\phi_3,\phi_4]$ 
and symmetrizes w.r.t.~the exchange of both pairs. Note that the outer parenthesis indicates 
the map defined in (\ref{4TENSORR}), and the inner parenthesis the map defined by (\ref{3BRACKET}). 
The fundamental identity is a direct consequence of the invariance of the 4-tensor under the Leibniz action 
and the fact that this action w.r.t.~$\pi(\varphi_1,\varphi_2)\in \frak{g}^*$ can be written in terms 
of the `3-bracket' (\ref{3BRACKET}) as 
 \be\label{twoargumentVAR}
  \delta_{\pi(\varphi_1,\varphi_2)}\phi \ = \ \{\varphi_1,\varphi_2,\phi\}\;. 
 \ee
Indeed, with this formula, varying the 4-tensor under $\delta_{\pi(\varphi_1,\varphi_2)}$ yields the right-hand side 
of (\ref{FUNDAMENTALIDD}), which gives zero by the above-mentioned invariance of the 4-tensor. 
To complete the proof it remains to prove (\ref{twoargumentVAR}). 
To this end we recall $\omega(\vartheta(\eta))=-\Theta(\omega,\eta)$ and compute 
 \be
 \begin{split}
  \langle \delta_{\pi(\varphi_1,\varphi_2)}\phi_1,\phi_2\rangle  \ &= \ \langle \rho_{\vartheta(\pi(\varphi_1,\varphi_2))}\phi_1,\phi_2\rangle \\
  \ &= \ -\pi(\phi_1,\phi_2)(\vartheta(\pi(\varphi_1,\varphi_2))) \\
  \ &= \ \Theta(\pi(\varphi_1,\varphi_2),\pi(\phi_1,\phi_2)) \\
  \ &= \ \{ \varphi_1,\varphi_2,\phi_1,\phi_2\} \\
  \ &= \ \langle \{\varphi_1,\varphi_2,\phi_1\}, \phi_2\rangle\;, 
 \end{split}
 \ee
using (\ref{3BRACKET}) in the last step.  Since this holds for arbitrary $\phi_2$ and the metric $\langle\,,\rangle$ 
is non-degenerate, (\ref{twoargumentVAR}) follows.

So far our discussion has been completely general, so that the resulting structures may be relevant for theories with less than 
maximal or no supersymmetry. We now turn to the special case that  the above 4-tensor is totally antisymmetric, as will be required for the ${\cal N}=8$ supersymmetry to be discussed momentarily, in which case there is an alternative form of the fundamental identity. 
 We then have 
 \be
 \begin{split}
  \{\{\varphi_1,\varphi_2,\phi_1\}, \phi_2, \phi_3, \phi_4\} \ &= \ 
  -\{\phi_2,\phi_3,\phi_4,\{\varphi_1,\varphi_2,\phi_1\}\} \\
  \ &= \ -\langle \{\phi_2,\phi_3,\phi_4\},\{\varphi_1,\varphi_2,\phi_1\}\rangle \\
  \ &= \ -\langle \{\varphi_1,\varphi_2,\phi_1\},\{\phi_2,\phi_3,\phi_4\} \rangle \\
  \ &= \ - \{\varphi_1,\varphi_2,\phi_1, \{\phi_2, \phi_3, \phi_4\} \}\;, 
 \end{split} 
 \ee
where we used the total antisymmetry in the first line,  the defintion 
(\ref{3BRACKET}) in the second and fourth line, and the symmetry of the metric in the third line. Under the sum $\sum_{{\rm sym}}$ that 
antisymmetrizes over the arguments $\phi_1,\ldots,\phi_4$ the left-hand side vanishes by (\ref{FUNDAMENTALIDD}). 
We thus obtain,  for a totally antisymmetric 4-tensors, the following form of the fundamental identity 
 \be\label{alternatefundamentalist} 
  \sum_{{\rm sym}}\{\varphi_1,\varphi_2,\phi_1, \{\phi_2, \phi_3, \phi_4\} \} \ = \ 0 \;, 
 \ee
where the sum antisymmetrizes over $\phi_1,\ldots,\phi_4$. 

\subsection{${\cal N}=8$ supersymmetry} 

We now define the complete ${\cal N}=8$ superconformal field theory Lagrangian and prove supersymmetric invariance. 
The matter fields, which are scalars and Majorana spinor fields under the $SO(1,2)$ Lorentz group (with signature $(-++)$), 
live in the representation $R$, with invariant metric $\langle\,,\rangle $. 
Moreover, the scalar fields $\phi^I$, $I=1,\ldots, 8$,  transform in the vector representation $8_v$ of the $SO(8)$ R-symmetry group, 
the spinors $\psi_A$, $A=1,\ldots, 8$,  transform 
in the spinor representation $8_s$, and the 
supersymmetry parameter $\epsilon_{\dot{A}}$, $\dot{A}=1,\ldots, 8$, 
in the conjugate spinor representation $8_c$. The $SO(8)$ gamma matrices are $\Gamma^{I}_{A\dot{A}}$
and their transpose $\bar{\Gamma}^I_{\dot{A}A}$, satisfying the familiar relations, and we define 
$\Gamma^{IJ}=\Gamma^{[I}\bar{\Gamma}^{J]}$, etc. 
The gamma matrices of the Lorentz group $SO(1,2)$, which commute with the $\Gamma^I$, are denoted by $\gamma^{\mu}$. 
Due to the spinor fields being Majorana, we have relations like
$\bar{\psi}\gamma^{\mu}\chi=-\bar{\chi}\gamma^{\mu}\psi$ and $\bar{\psi}\chi=\chi\bar{\psi}$ 
and hence $\bar{\psi}\,\Gamma^{I}\epsilon=\bar{\epsilon}\,\bar{\Gamma}^I\psi$. Note that here and in the following we will usually 
suppress the $SO(8)$ spinor indices, which can always be restored by recalling the $SO(8)$ representation of the spinors involved, 
but we will display the vector indices. 
In general, we follow the spinor conventions of \cite{Bandos:2008jv}. 

We begin by giving the free (ungauged) Lagrangian for the scalar and spinor fields, 
 \be\label{freeLagrangian}
  {\cal L}_0 \ = \ -\tfrac{1}{2}\langle \partial^{\mu}\phi^I, \partial_{\mu}\phi^{I}\rangle 
  -\tfrac{i}{2}\langle \bar{\psi},\gamma^{\mu}\partial_{\mu}\psi\rangle\;, 
 \ee
which is invariant, up to total derivatives, under the  ${\cal N}=8$ supersymmetry transformations 
 \be
  \delta_{\epsilon}\phi^I \ = \ i\bar{\psi}\Gamma^{I}\epsilon\;, \qquad
  \delta_{\epsilon}\psi \ = \ \gamma^{\mu}\Gamma^{I} \partial_{\mu}\phi^I\epsilon\;. 
 \ee
This follows by a quick computation, using that the spinorial supersymmetry parameter $\epsilon$ and 
the $\Gamma^I$ are just numbers w.r.t.~the representation $R$ and hence can be freely moved in and out of the inner product.  
The gauged theory is now obtained by introducing gauge vectors $A_{\mu}\in \frak{g}^*$ and promoting the partial derivatives to covariant derivatives, 
 \be
   D_{\mu} \ =\  \partial_{\mu}-\rho_{\vartheta(A_{\mu})}\;. 
 \ee
These covariant derivatives  satisfy   
  \be
    [D_{\mu}, D_{\nu}] \ = \ -\rho_{\vartheta(F_{\mu\nu})}\;, 
   \ee
 with the field strength defined in (\ref{fieldSTRENGTH}).  Due to this non-commutativity, 
 the action is no longer supersymmetric: a field strength term is generated. This term can be cancelled by adding a 
 Chern-Simons term, which varies as (\ref{generalCSvar}), and assigning appropriate supersymmetry transformations to 
 the gauge vectors. These then have to be varied inside covariant derivatives, whose cancellation in turn requires 
 higher-order couplings in the scalars, which can be efficiently written in terms of the 4-tensor (\ref{4TENSORR}) and the 3-bracket (\ref{3BRACKET}). 
 The total Lagrangian reads  
 \be\label{totalLagrangian}
  \begin{split}
   {\cal L} \ = \ &-\tfrac{1}{2}\langle D^{\mu}\phi^I, D_{\mu}\phi^I\rangle -\tfrac{i}{2}\langle \bar{\psi},\gamma^{\mu}D_{\mu}\psi\rangle
  +\tfrac{1}{2}{\cal L}_{\rm CS}\\
   &-\tfrac{i}{4} \{\phi^I, \phi^J, \bar{\psi}, \Gamma^{IJ}\psi\} 
   -\tfrac{1}{12}\langle \{\phi^I, \phi^J, \phi^K\}, \{\phi^I, \phi^J, \phi^K\}\rangle \;, 
  \end{split}
 \ee  
with supersymmetry rules
 \be\label{SUSYrules}
  \begin{split}
   \delta_{\epsilon}\phi^I \ &= \ i\bar{\psi}\,\Gamma^{I}\epsilon\;, \\
  \delta_{\epsilon}\psi \ &= \ \gamma^{\mu}\,\Gamma^{I} D_{\mu}\phi^I\epsilon + \tfrac{1}{6}\{\phi^I, \phi^J, \phi^K\}\Gamma^{IJK}\epsilon\;, \\
  \delta_{\epsilon}A_{\mu} \ &= \ -i \bar{\epsilon}\,\gamma_{\mu}\,\bar{\Gamma}^{I}\pi(\psi, \phi^I)\;.  
  \end{split}
 \ee 
Note that the supersymmetry rules for $A_{\mu}\in \frak{g}^*$ are naturally written in terms of the 
map $\pi: R \wedge R\rightarrow \frak{g}^*$ defined in (\ref{BestPI}). 
 
In the remainder of this section we will show that the proof of supersymmetric invariance can be entirely formulated 
in terms of the invariantly defined maps of the previous subsection, so that the above theory in particular is valid for 
infinite-dimensional gauge groups (subject to the quadratic and linear constraint). 
We first consider the variation of the fermion kinetic term, which gives a new contribution due to the 
non-commutativity of covariant derivatives: 
 \be\label{LpsiVAR}
 \begin{split}
  \delta_{\epsilon}{\cal L}_{\psi} \ &= \ 
  \tfrac{i}{2}\bar{\epsilon}\,\gamma^{\mu\nu}\,\bar{\Gamma}^I\langle \psi, [D_{\mu},D_{\nu}]\phi^I\rangle \\
  \ &= \ -\tfrac{i}{2}\bar{\epsilon}\,\gamma^{\mu\nu}\,\bar{\Gamma}^I\langle \psi, \rho_{\vartheta(F_{\mu\nu})}\phi^I\rangle \\
  \ &= \ -\tfrac{i}{2}\bar{\epsilon}\,\gamma^{\mu\nu}\,\bar{\Gamma}^I (\pi(\psi, \phi^I))(\vartheta(F_{\mu\nu}))\;, 
 \end{split}
 \ee 
recognizing the map (\ref{BestPI}) in the last step. 
This is cancelled by the Chern-Simons term that varies according to (\ref{generalCSvar}) as 
 \be
 \begin{split}
  \delta_{\epsilon}{\cal L}_{\rm CS} \ &= \ \varepsilon^{\mu\nu\rho}\, \Theta(\delta_{\epsilon} A_{\mu}, F_{\nu\rho}) \\
   \ &= \ -\varepsilon^{\mu\nu\rho} (\delta_{\epsilon}A_{\mu}) (\vartheta(F_{\nu\rho}))\\
  \ &= \ \varepsilon^{\mu\nu\rho} i \bar{\epsilon}\,\gamma_{\mu}\,\bar{\Gamma}^{I}\pi(\psi, \phi^I)(\vartheta(F_{\nu\rho}))\;. 
 \end{split}
 \ee
This cancels (\ref{LpsiVAR}), provided we pick a pre-factor of $\frac{1}{2}$ for the Chern-Simons term and use conventions with 
$\gamma^{\mu\nu}=\varepsilon^{\mu\nu\rho}\gamma_{\rho}$. 
Next, the supersymmetry variation of the gauge field inside the scalar kinetic term produces a $\phi^3\psi$ contribution that needs to be cancelled 
from the supersymmetry variation of the fermion kinetic term due to the extra term in $\delta\psi$ and through the lowest-order  variation of the 
Yukawa coupling. The variation of the scalar kinetic term reads 
 \be\label{VARsclaar}
 \begin{split}
  \delta_{\epsilon}S_{\phi} \ &= \ \langle D^{\mu}\phi^I, \rho_{\vartheta(\delta_{\epsilon}A_{\mu})}\phi^I \rangle 
  \ = \ (\pi(D^{\mu}\phi^I, \phi^I))(\vartheta(\delta_{\epsilon}A_{\mu})) \\
  \ &= \ \Theta(\delta_{\epsilon}A_{\mu}, \pi(\phi^I, D^{\mu}\phi^I))\\
  \ &= \ -i\bar{\epsilon}\gamma_{\mu}\bar{\Gamma}^J \Theta(\pi(\psi, \phi^J), \pi(\phi^I, D^{\mu}\phi^I)) \\
  \ &= \ -i\bar{\epsilon}\gamma_{\mu}\bar{\Gamma}^J \{\psi, \phi^J, \phi^I, D^{\mu}\phi^I\}\;, 
 \end{split}
 \ee
while the variation of the fermion kinetic term becomes 
 \be\label{FERMKINVARRR}
  \begin{split}
    \delta_{\epsilon}S_{\psi} \ &= \ -\tfrac{i}{6}\bar{\epsilon}\,\bar{\Gamma}^{IJK}\langle \{\phi^I,\phi^J, \phi^K\},\gamma^{\mu} D_{\mu}\psi\rangle\\
    \ &= \ -\tfrac{i}{6}\bar{\epsilon}\,\bar{\Gamma}^{IJK} \{\phi^I, \phi^J,\phi^K,  \gamma^{\mu} D_{\mu}\psi\}\\
    \ &= \ \tfrac{i}{2}\{\phi^I, \phi^J,\bar{\psi}, \gamma^{\mu}D_{\mu}\phi^K\} \Gamma^{IJK}\epsilon \;, 
  \end{split}
 \ee   
where we integrated by parts and used the antisymmetry of $\{\cdot,\cdot,\cdot,\cdot\}$ in its last two arguments.  
Finally, we have to compute the variation of the Yukawa couplings proportional to $\phi^2\bar{\psi}\psi$ under the variation 
lowest order in scalars: 
 \be\label{YKvar}
 \begin{split}
  \delta_{\epsilon}S_{\rm Yukawa} \ &= \ -\tfrac{i}{2} \{\phi^I, \phi^J, \bar{\psi}, \Gamma^{IJ}\Gamma^{K}\gamma^{\mu} D_{\mu}\phi^K\}\epsilon \\
  \ &= \ -\tfrac{i}{2} \{\phi^I, \phi^J, \bar{\psi}, \Gamma^{IJK}\gamma^{\mu} D_{\mu}\phi^K\}\epsilon 
  + i \bar{\epsilon}\gamma^{\mu}\bar{\Gamma}^J \{\psi, \phi^J, \phi^I,D_{\mu}\phi^I\}\;. 
 \end{split}
 \ee
 Here we used the $SO(8)$ gamma matrix identity 
 \be
  \Gamma^{IJ}\Gamma^K \ = \ \Gamma^{IJK}-2\,\delta^{K[I}\Gamma^{J]}\;, 
 \ee
and the total antisymmetry of the 4-tensor. 
The first term on the right-hand side of (\ref{YKvar}) cancels against (\ref{FERMKINVARRR}), and the second term on the 
right-hand side of (\ref{YKvar}) cancels against (\ref{VARsclaar}). Note, in particular, that this cancellation required 
the total antisymmetry of the 4-tensor, which is the first time this assumption is needed, and hence this linear constraint 
is required by supersymmetry. 
 
Finally, we turn to  the terms of the structural form $\phi^5\psi$. One source for such terms is the variation of the Yukawa couplings 
under the term in $\delta_{\epsilon}\psi$ cubic in $\phi$. Using the gamma matrix identity 
 \be
  \Gamma^{IJ}\Gamma^{KLP} \ = \ \Gamma^{IJKLP}-6\,\delta^{[\underline{I}[K}\Gamma^{LP]\underline{J}]}
  -6\, \delta^{I[K}\delta^{|J|L}\Gamma^{P]}\;, 
 \ee
this gives three contributions: proportional to $\Gamma^5$, $\Gamma^3$ and $\Gamma^1$. The first two vanish as a a consequence 
of the fundamental identity in the form (\ref{alternatefundamentalist}). The final one proportional to $\Gamma^1$ yields 
 \be
  \delta_{\epsilon}S_{\rm Yukawa} \ = \ \tfrac{i}{2}\{\phi^I,\phi^J, \bar{\psi},\{\phi^I, \phi^J, \phi^K\}\}\Gamma^K \epsilon\;. 
 \ee
This precisely cancels against the supersymmetry variation of the scalar potential.

\newpage

\section{SDiff$_3$ superconformal field theory}

In this section we introduce the ${\cal N}=8$ superconformal field theories based on the Lie algebra 
of (infinitesimal) volume-preserving diffeomorphisms on a 3-manifold. In the first subsection we show that 
there is natural embedding tensor on the dual space of one-forms that satisfies the quadratic constraint. 
The space of one-forms  thus forms  a Leibniz-Loday algebra  and carries an associated L$_{\infty}$ structure, 
realizing the axioms of a (generalized) Courant algebroid, 
which will be discussed  in the second subsection. The complete theory will be given in the final subsection, 
where we show that it is equivalent to the Bandos-Townsend theory.

\subsection{Volume-preserving diffeomorphisms} 

We consider the group SDiff$_3$ of 
volume preserving diffeomorphisms on a 3-manifold carrying a volume-form but no further a priori structure. 
In particular, we do not use a metric,  because a metric cannot be SDiff$_3$ invariant. 
The volume form gives rise to invariant tensors 
 \be\label{volumeforms}
 \epsilon_{ijk} \ \equiv \ e \varepsilon_{ijk} \;, \qquad \epsilon^{ijk} \ \equiv \ e^{-1} \varepsilon^{ijk}\;, 
 \ee
where $\varepsilon$ is the constant Levi-Civita symbol with $\varepsilon_{123}=\varepsilon^{123}=1$, 
and $e(y)$ a function on the 3-manifold $M_3$ with coordinates 
$y^i$, $i=1,2,3$, measuring the volume. 
The Lie algebra $\frak{sdiff}_3$ of the group SDiff$_3$ is generated by vector fields $v^{i}(y)$ having zero divergence: 
 \be\label{divzero}
  \partial_i(ev^{i}) \ = \ 0\;. 
 \ee
The Lie algebra structure on $\frak{sdiff}_3$ is given by the familiar Lie bracket of vector fields, 
 \be\label{Liebracket}
  [v,w]^i \ = \ v^j\partial_jw^i - w^j\partial_jv^i\;,  
 \ee
which preserves the zero-divergence condition (\ref{divzero}). 
Natural representations of this Lie algebra include all actions on tensor fields generated 
by Lie derivatives, as for instance on functions $f$ or one-forms $\omega=\omega_i{\rm d}y^i$: 
 \be\label{LiederREP}
  {\cal L}_{v}f \ = \ v^i\partial_if\;, \qquad {\cal L}_{v}\omega_{i} \ = \ v^{j}\partial_j\omega_i+\partial_iv^j \omega_{j}\;, 
 \ee
which close according to (\ref{Liebracket}).

The Lie algebra of volume-preserving diffeomorphisms of an arbitrary $n$-manifold does not carry an invariant 
metric. Indeed, the naive candidate $\langle v, w\rangle = \int {\rm d}^ny \,e \,g_{ij} v^i w^j$ obtained by picking 
a metric $g_{ij}$ for which $\sqrt{\det{g}}$ equals the given volume element, is not invariant under all volume-preserving 
diffeomorphisms but only under its (necessarily finite-dimensional) isometry (sub-)group.   
Since in order to write a conventional Yang-Mills or Chern-Simons gauge theory one needs a Lie algebra 
carrying an invariant bilinear form  it follows that one cannot formulate a gauge theory based on the volume-preserving 
diffeomorphisms of an arbitrary $n$-manifold.  3-manifolds are special in that, under certain topological assumptions, 
there is an invariant bilinear form, as will be reviewed below, but this bilinear form is non-local. 
A manifestly local formulation, without any topological assumptions,  can instead be obtained by transporting the Lie algebra structure on vector 
fields to a Leibniz-Loday algebra (and thus an L$_{\infty}$ structure) on the dual one-forms. 

We thus consider the space $\frak{g}^*$ of one-forms dual to the Lie algebra $\frak{g}\equiv \frak{sdiff}_3$ defined above. 
The pairing between vectors $v$ and covectors $\omega$ is given by 
 \be\label{covectorvector}
  \omega(v) \ = \  \int {\rm d}^3y\,e\, \iota_{v}\omega\;, 
 \ee
where $\iota_{v}\omega=v^i\omega_i$, and the coadjoint representation is given by Lie derivatives (\ref{LiederREP}) 
acting on one-forms, under which this pairing is manifestly invariant. 
Let us emphasize that here we use a more extensive notion of `dual space' in that the above pairing is degenerate: it vanishes 
for any exact one-form, ${\rm d}f(v)=0$, as follows by integration by parts and (\ref{divzero}).
Accordingly, in the literature the dual space to $\frak{g}\equiv \frak{sdiff}_3$ is more conventionally identified with the quotient space 
of one-forms modulo 
 exact one-forms \cite{ArnoldHydro}.  However, in the present context it is convenient to stay on the space of all one-forms 
and to interpret the `modding out of exact forms' as a
gauge redundancy, which indeed emerges naturally in the formulation below. 
By a slight abuse of the usual nomenclature, we continue to refer to the space of one-forms as the dual space.

We next have to define an embedding tensor $\vartheta:\frak{g}^*\rightarrow \frak{g}$. Given the volume form (\ref{volumeforms}) 
there is a natural such map given, for $\eta\in\frak{g}^*$, by 
 \be\label{firstembedding}
  \vartheta(\eta)^i \ \equiv \ \epsilon^{ijk}\partial_j\eta_k\;. 
 \ee
Clearly, $\vartheta(\eta)\in \frak{g}$ since 
$\partial_i(e\vartheta(\eta)^i)  =  \varepsilon^{ijk}\partial_i\partial_j\eta_k=0$. 
This map gives rise to an associated bilinear form, 
 \be\label{embeddingAGAIN}
  \Theta(\omega, \eta) \ \equiv \ -\omega(\vartheta(\eta)) \ = \ -\int  \omega\wedge {\rm d}\eta  \;, 
 \ee
which is a natural diffeomorphism invariant bilinear form on one-forms. 
Note that this bilinear form is degenerate in that 
 \be
  \Theta({\rm d}f, \eta)  \ = \ 0 \quad \forall \eta\;, 
 \ee
in agreement with the fact that there is no isomorphism between $\frak{g}$ and $\frak{g}^*$ and that the Lie algebra 
structure on $\frak{g}$ cannot be transported to $\frak{g}^*$.

We are now ready to define the Leibniz algebra as above by means of the coadjoint action, 
which is here given by Lie derivatives, 
 \be\label{oneThetaDEF}
 \omega\circ \eta \ \equiv \ {\cal L}_{\vartheta(\omega)}\eta
  \;. 
 \ee
This relation implies that, upon pairing with a vector $v$, 
 \be
  (\omega\circ \eta)(v) \ = \ \Theta(\omega, {\cal L}_{v}\eta)\;. 
 \ee
To verify this we start from the right-hand side and  
use the manifest covariance of $\vartheta$ and the invariance of the pairing, 
 \be
 \begin{split}
  \Theta(\omega, {\cal L}_{v}\eta) \ &= \ -\omega(\vartheta({\cal L}_v\eta)) \ = \ -\omega({\cal L}_v(\vartheta(\eta)))
  \ = \ \omega({\cal L}_{\vartheta(\eta)}v) \\
  \ &= \ -({\cal L}_{\vartheta(\eta)}\omega)(v) 
  \ = \ -(\eta\circ \omega)(v) \ = \ (\omega\circ \eta)(v)\;,  
 \end{split}
 \ee
where the last equality follows from the pairing being zero for the symmetric, trivial part of the Leibniz product.

We next have to verify that the quadratic constraint (\ref{quadconstr}) is satisfied. 
To this end it is convenient to write the Leibniz product (\ref{oneThetaDEF})  more explicitly, 
using some invariant notation. Given the volume form, there is a canonical map $\star: \frak{g}\rightarrow\Omega_2$
from divergence-free vector fields to (closed) 2-forms, in local coordinates given by 
 \be\label{closed2form}
  (\star \,v)_{ij} \ \equiv \ \epsilon_{ijk} v^k\;. 
 \ee
The embedding tensor (\ref{firstembedding}) then satisfies 
 \be\label{dinversetheta}
  \star \, \vartheta(\omega) \ = \ {\rm d}\omega\;. 
 \ee
There is also a canonical map $\times :\frak{g}\wedge \frak{g}\rightarrow \frak{g}^*$ defined by 
 \be\label{crossproduct}
  v\times w \ \equiv \ -\iota_{v}(\star\,w) \qquad \Rightarrow \qquad (v\times w)_i \ = \ \epsilon_{ijk}\, v^j w^k\;, 
 \ee 
where we displayed the expression in local coordinates.  
This `cross product' is mapped under $\vartheta$ to (minus) the Lie bracket on $\frak{g}$, 
 \be\label{crossotoLIE}
  \vartheta(v\times w) \ = \ -[v,w]\;, 
 \ee
as follows by a quick computation using the zero-divergence conditions (\ref{divzero}). 
A consequence of this relation and the symmetry of $\Theta$ is that for any vectors $v, w$ and covector $\omega$
 \be\label{NICerelation}
  (v\times w)(\vartheta(\omega)) \ = \ -\Theta(\omega, v\times w) \ = \ \omega(\vartheta(v\times w)) \ = \ -\omega([v,w])\;. 
 \ee

Using the above formulae and `Cartan's magic identity' for Lie derivatives acting on forms, ${\cal L}_{v}=\iota_{v}\,{\rm d}+{\rm d}\, \iota_v$, 
we then compute 
 \be 
  \begin{split}
   \omega \circ \eta \ = \ {\cal L}_{\vartheta(\omega)}\eta \ = \ \iota_{\vartheta(\omega)}({\rm d}\eta)+{\rm d}(\iota_{\vartheta(\omega)}\eta)
   \ = \ \iota_{\vartheta(\omega)}(\star\,\vartheta(\eta))+{\rm d}(\iota_{\vartheta(\omega)}\eta)\;, 
  \end{split}
 \ee  
and thus by means of  (\ref{crossproduct}) 
  \be\label{symmRELLL2}
  \omega\circ \eta \ = \ - \vartheta(\omega)\times \vartheta(\eta)  \ + \ {\rm d}(\iota_{\vartheta(\omega)}\eta)\;. 
 \ee
This formula is intriguing in that the first term, which is manifestly antisymmetric, looks like the $\frak{so}(3)$ Lie algebra, 
but of course differs from it in two important respects: first, in absence of a metric, 
the product $\times$ does not map vectors to vectors but rather to one-forms; 
second, the insertions of the map $\vartheta$ are needed. 
Consequently, the first term by itself does not define a Lie algebra. 
However, acting with $\vartheta$ on (\ref{symmRELLL2}) and using (\ref{crossotoLIE}) one learns that the quadratic constraint 
is satisfied, 
 \be\label{quadCONSTR123345}
  \vartheta(\omega\circ \eta) \ = \ \big[\vartheta(\omega), \vartheta(\eta)\big]\;,  
 \ee 
and hence, by the general arguments of sec.~2, (\ref{symmRELLL2}) defines a Leibniz-Loday algebra.

\subsection{L$_{\infty}$ algebra and Courant algebroid}

In order to further elucidate the above algebraic structure on the space $\frak{g}^*$ of one-forms, 
we will now discuss its associated L$_{\infty}$ structure and show that it provides a novel realization 
of a Courant algebroid. 
First, let us display the  symmetric and antisymmetric parts (\ref{symmantisymm}), 
which can be read off from (\ref{symmRELLL2}): 
 \be
  \begin{split}
   [\omega,\eta] \ &= \  - \vartheta(\omega)\times \vartheta(\eta)  
   +\tfrac{1}{2} {\rm d}(\iota_{\vartheta(\omega)}\eta-\iota_{\vartheta(\eta)}\omega) \;, \\
   \{\omega, \eta\} \ &= \ \tfrac{1}{2} {\rm d}(\omega\bullet \eta)\;, 
  \end{split}
 \ee 
with the symmetric pairing 
 \be\label{symmetricbullet}
  \omega\bullet \eta \ \equiv \ \iota_{\vartheta(\omega)}\eta+ \iota_{\vartheta(\eta)}\omega \;.
 \ee
We thus identified the pairing implicitly defined in (\ref{DREL}) for the symmetric part of the product, 
where ${\cal D}$ is given by the de Rham differential ${\rm d}$.  
 By explicit computation one finds relations for Leibniz products in which one factor is exact: 
 \be\label{LeibniztrivFac}
  \begin{split}
   \omega \circ {\rm d}f \ &= \ {\rm d}(\omega\bullet {\rm d}f)\;, \\
   {\rm d}f \circ \omega \ &= \ 0\;. 
  \end{split}
 \ee  
Moreover, with the above formulas one may compute the failure of the bracket to satisfy the Jacobi identity (the `Jacobiator'): 
 \be\label{JacobiatorRRR}
  {\rm Jac}(\omega_1, \omega_2, \omega_3) \ \equiv \ 3[[\omega_{[1}, \omega_{2}], \omega_{3]}] \ = \ 
 \tfrac{1}{2}{\rm d}\big([\omega_{[1}, \omega_{2}^{}]\bullet \omega_{3]}\big)\;. 
 \ee

Although the Jacobi identity is not satisfied, and hence the bracket $[\cdot,\cdot]$ does not define a Lie algebra, 
this algebraic structure defines a strongly homotopy Lie algebra or L$_{\infty}$ algebra, as we will show in the following. 
A L$_{\infty}$ algebra is defined on a graded vector space $X$, which we here take to be 
 \be
  X \ = \ X_1+X_0\,, 
 \ee
where the subscripts denote the grading, and $X_0=\frak{g}^*$ is the space of one-forms on which 
the Leibniz algebra is defined. The space of degree one will be taken to be the space of smooth functions (zero-forms) on 
the 3-manifold, $X_1={\cal C}^{\infty}(M_3)$. In order to specify the L$_{\infty}$ structure on this space, we have 
to define multilinear, graded antisymmetric maps or brackets $\ell_i$, where  $i=1,2,3,\ldots$ indicates the number of arguments, 
of intrinsic degree $i-2$, 
satisfying certain quadratic generalized Jacobi identities.  
The only non-trivial $\ell_1$ maps $X_1$ to $X_0$ and is given by $\ell_1={\rm d}$. 
The map $\ell_2$ is given, for both arguments in $X_0$, by the above bracket, 
while for one argument $f\in X_1$ and one argument $\omega\in X_0$ it can be written in terms of the bilinear operation 
(\ref{symmetricbullet}), 
 \be\label{ultimate2Brackets}
  \ell_2(\omega, \eta) \ = \ [\omega, \eta]\;, \qquad 
  \ell_2(\omega,f) \ = \ \,\tfrac{1}{2}\,\omega \bullet {\rm d}f\;. 
 \ee
There are no further 2-brackets compatible with the grading. The second relation in here is fixed by the 
L$_{\infty}$ relation stating that $\ell_1$ acts as a derivation on $\ell_2(\cdot, \cdot)$, 
 \be
  \ell_1(\ell_2(f,\omega)) \ = \ \ell_2(\ell_1(f),\omega)+(-1)^{|f|}\ell_2(f, \ell_1(\omega)) \ = \ \ell_2(\ell_1(f),\omega)\;, 
 \ee
where the second equality holds because of $\ell_1$ acting trivially on $X_0$.  
Using (\ref{LeibniztrivFac}) we may now verify that this L$_{\infty}$ relation is satisfied for (\ref{ultimate2Brackets}). 
Finally, the 3-bracket is determined from the L$_{\infty}$ relation
 \be
  {\rm Jac}(\omega_1,\omega_2,\omega_3) + \ell_1(\ell_3(\omega_1,\omega_2,\omega_3)) \ = \ 0\;. 
 \ee
Comparing with (\ref{JacobiatorRRR}) we thus infer
  \be
  \ell_3(\omega_1, \omega_2, \omega_3) \ = \ -\tfrac{1}{2}[\omega_{[1}, \omega^{}_{2}]\bullet \omega_{3]}\;. 
 \ee

The claim is that the above generalized brackets $\ell_i$, $i=1,2,3$, define an L$_{\infty}$ algebra, 
with no higher brackets needed. 
Instead of discussing the explicit verification of this claim  in more detail, in the following we show that 
this algebraic  structure on one-forms defines (a mild generalization of) a Courant algebroid, so that it follows from 
the theorem in \cite{roytenberg-weinstein} (or the more general results in \cite{Hohm:2017cey}) 
that this defines an L$_{\infty}$ algebra. 

We begin by recalling the definition  of a Courant algebroid, using the `alternative' definition given in \cite{roytenberg}. 
A Courant algebroid is a vector bundle $E\rightarrow M$ equipped with a 
non-degenerate symmetric bilinear form $(\,,)$,  a bilinear (not necessarily antisymmetric) 
operation $\circ $ on $\Gamma(E)$, 
and an anchor map $\rho\,: E\; \rightarrow\; TM$, so that the following axioms hold for $x,y,z\in \Gamma(E)$: 
 \be
  \begin{split}
   (i)&\;\;\text{$\Gamma(E)$ is a Leibniz algebra w.r.t.}\;\, \circ \\
   (ii)&\;\; \rho(x\circ y) \ = \ [\rho(x),\rho(y)]\\
   (iii)&\;\; x\circ fy \ = \ f(x\circ y) + (\rho(x)f)y\;, \qquad f \ \in \ {\cal C}^{\infty}(M) \\
   (iv)&\;\; x\circ x \ = \ \tfrac{1}{2} D(x,x) \\
   (v)&\;\; \rho(x)(y,z) \ = \ (x\circ y,z) + (y,x\circ z)
  \end{split}
 \ee 
where $D$ is implicitly defined by 
 \be\label{DDEF}
  (Df, x) \ = \ \rho(x)f\;. 
 \ee 

For our example we can take $E=T^*M$ and $\rho=\vartheta$. The quadratic constraint (\ref{quadCONSTR123345})
is then equivalent to $(ii)$ and implies the Leibniz algebra relations $(i)$. The relation $(iii)$ follows by a quick computation 
from the original definition (\ref{oneThetaDEF}) of the Leibniz algebra. To satisfy $(iv)$ we have to identify $D={\rm d}$ and take the metric to be 
 \be
  (\omega, \eta) \ \equiv \ \omega \bullet \eta 
  \;, 
 \ee
for which  (\ref{DDEF}) holds. Finally, $(v)$ follows from the fact that the symmetric pairing is manifestly covariant 
under the action of the Lie derivative w.r.t.~$\vartheta(\omega)$. 
This completes the proof that for an arbitrary 3-manifold $M$ with a volume form, the cotangent bundle $E=T^*M$ 
together with the Leibniz product defined above and the anchor map given by the embedding tensor (\ref{firstembedding})
satisfies the above axioms. It then follows with the theorem in \cite{roytenberg-weinstein} 
that the cotangent bundle naturally carries an L$_{\infty}$ structure with the highest bracket being a 3-bracket. 
More precisely, the above structure defines a mild generalization of a Courant algebroid in that the definition 
usually takes the anchor map to be a bundle map, which means that it is compatible with the projection onto each fibre. 
Since the embedding tensor map (\ref{firstembedding}) involves the derivative of a one-form, it does not define a 
bundle map,\footnote{We thank Alan Weinstein for pointing this out to us.} 
but it may be verified that the proof in \cite{roytenberg-weinstein} does not actually depend on this assumption.

\subsection{Bandos-Townsend theories}

We will now investigate the ${\cal N}=8$ superconformal field theory defined in sec.~2 
for the above embedding tensor (\ref{firstembedding}), (\ref{embeddingAGAIN}) and show that 
the resulting theory is equivalent to the Bandos-Townsend theory constructed in \cite{Bandos:2008jv} --- at least under 
favorable topological assumptions.  We begin by inspecting the topological sector given by the 
Leibniz-Chern-Simons theory based on a gauge vector taking values in the space $\frak{g}^*$ of one-forms. 
Thus, the gauge fields are `one-form-valued' one-forms, 
 \be
  A(x,y) \ = \ A_{\mu i}(x,y){\rm d}x^{\mu}\otimes {\rm d}y^i\;. 
 \ee
The gauge transformations (\ref{FullDeltaA}) then read 
 \be\label{gaugeAform}
  \delta A_{\mu i} \ = \ \partial_{\mu}\lambda_i -(A_{\mu}\circ \lambda)_i - \partial_i\lambda_{\mu}\;, 
 \ee 
where we recalled, from the previous subsection, that ${\cal D}={\rm d}$ in terms of the internal de Rham differential. 
The gauge symmetry parameterized by $\lambda_{\mu}$ implements the redundancy that one-forms related by 
exact one-forms are physically indistinguishable, as alluded to in the paragraph containing (\ref{covectorvector}). 
Employing the L$_{\infty}$ structures discussed above, (\ref{gaugeAform}) can also be written in terms of $\lambda\in X_0$, $\lambda_{\mu}\in X_1$
as 
 \be
  \delta A_{\mu} \ = \ \partial_{\mu}\lambda \ - \ \ell_2(A_{\mu},\lambda) \ - \ \ell_1(\tilde{\lambda}_{\mu})\;, 
 \ee
where we defined $\tilde{\lambda}_{\mu}=\lambda_{\mu}+\tfrac{1}{2}A_{\mu}\bullet \lambda \in X_1$.  

In order to relate to the formulation in \cite{Bandos:2008jv} we write for the image of the gauge fields 
under the embedding tensor map $\vartheta$, 
 \be\label{sAREL}
  s_{\mu}{}^{i} \ \equiv \ \vartheta(A_{\mu})^i \ = \ \epsilon^{ijk}\partial_j A_{\mu k}\;. 
 \ee
Acting with $\vartheta$ on (\ref{gaugeAform}) and writing $\xi=\vartheta(\lambda)$ we obtain with the quadratic constraint 
(\ref{quadCONSTR123345}) 
 \be
  \delta s_{\mu} \ = \ \partial_{\mu}\xi \ - \ [s_{\mu},\,\xi ]\;. 
 \ee
In particular, $[\cdot,\cdot]$ is now a genuine Lie bracket and the additional one-form redundancy with parameter $\lambda_{\mu}$ has disappeared, 
so this is a conventional Yang-Mills gauge transformation. 
Indeed, the formulation in \cite{Bandos:2008jv} starts with the Lie algebra $\frak{g}=\frak{sdiff}_3$ and introduces 
Yang-Mills gauge fields $s_{\mu}$ taking values in $\frak{g}$, from which 
the gauge fields $A_{\mu}$ are derived  via (\ref{sAREL}), assuming that the corresponding de Rham cohomology is 
trivial. The present formulation circumvents the need for such topological assumptions by viewing the $A_{\mu}$ 
as the fundamental fields, which requires taking the underlying gauge algebra to be a Leibniz algebra 
rather than a Lie algebra. 

Let us now verify that with the above identifications the Chern-Simons action (\ref{CSSS}), 
 \be\label{secondorthirdCS}
  S_{\rm CS} \ =  \ \int {\rm d}^3x\, \varepsilon^{\mu\nu\rho}\,\Theta\big(A_{\mu}, \partial_{\nu}A_{\rho}
  -\tfrac{1}{3}\,  A_{\nu}\circ A_{\rho}\big)\;, 
 \ee
coincides with the Chern-Simons-like action given in \cite{Bandos:2008jv}. 
We first consider the quadratic term, which can be rewritten by use of (\ref{embeddingAGAIN}) as 
 \be
  \Theta(A_{\mu}, \partial_{\nu}A_{\rho}) \ = \ -A_{\mu}(\vartheta(\partial_{\nu}A_{\rho}))
  \ = \ -A_{\mu}(\partial_{\nu}\vartheta(A_{\rho})) \ = \ -A_{\mu } (\partial_{\nu} s_{\rho})\;, 
 \ee
where we used that the external derivatives commute with the internal derivatives defining $\frak{sdiff}_3$ 
(note that the volume measure $e$ continues to depend only on $y$). 
Thus, in form notation we have 
 \be
  \int {\rm d}^3x\, \varepsilon^{\mu\nu\rho}\,\Theta\big(A_{\mu}, \partial_{\nu}A_{\rho}) \ = \ - \int {\rm d}^3 y\,e \int A_{i}\wedge {\rm d} s^{i}
   \ = \ - \int {\rm d}^3 y\,e\int {\rm d}s^i\wedge A_{i} \;. 
 \ee 
For the cubic term we compute with the quadratic constraint (\ref{quadCONSTR123345}) 
 \be
 \begin{split}
  \Theta(A_{\mu},A_{\nu}\circ A_{\rho}) \ &= \ -A_{\mu}(\vartheta(A_{\nu}\circ A_{\rho})) \ = \ -A_{\mu}([\vartheta(A_{\mu}), \vartheta(A_{\rho})])
  \ = \ -A_{\mu}([s_{\nu}, s_{\rho}])\\
  \ &= \ (s_{\nu}\times s_{\rho})(\vartheta(A_{\mu})) \ = \ (s_{\nu}\times s_{\rho})(s_{\mu})\\
  \ &= \ \int {\rm d}^3 y\,e\, \epsilon_{ijk}\, s_{\mu}{}^{i} s_{\nu}{}^{j} s_{\rho}{}^{k}\;, 
 \end{split} 
 \ee
where we used (\ref{NICerelation}) in the second line. Thus, in form notation w.r.t.~to the external space, 
the total Chern-Simons action (\ref{secondorthirdCS}) reads 
 \be\label{finalBTCS}
  S_{\rm CS} \ = \ -\int {\rm d}^3 y\,e \left(\int {\rm d}s^i\wedge A_{i} \ + \ \frac{1}{3}\, \epsilon_{ijk}\, s^{i}\wedge s^{j}\wedge s^{k}\right)\;, 
 \ee
which agrees with the Chern-Simons-like action given in  \cite{Bandos:2008jv}, see their eq.~(3.3).\footnote{More precisely, 
the sign of the first term differs, but this is field redefinition equivalent because we can send $A\rightarrow -A$ but keep $s$ unchanged, i.e., 
define $s=-\vartheta(A)$ in terms of the new $A$.} 

Let us note in passing that under the same topological assumptions that allowed for the above match with the 
Chern-Simons-like action of Bandos-Townsend, the latter action actually also has a conventional Chern-Simons interpretation, 
provided we accept a non-local bilinear form. (See also \cite{deAzcarraga:2010mr} for a related  discussion.)
To see this we recall from (\ref{closed2form}) that the volume form associates 
to any divergence-free vector field $v$ a closed 2-form $\star\, v$. Assuming again that the corresponding de Rham cohomology is 
trivial, $\star\, v$ is exact and hence there is a one-form $\omega_v$ so that 
 \be\label{newnot}
  \star\, v \ = \ {\rm d}\omega_v \;. 
 \ee
We then have the bilinear invariant defined as the integral over a natural 3-form,\footnote{This bilinear form 
is known as the \textit{helicity} or \textit{Hopf} invariant and arises, for instance, in hydrodynamics \cite{ArnoldHydro}.} 
 \be\label{invariantmetric}
  \langle v, w\rangle \ \equiv \ \int (\star\, v)\wedge {\rm d}^{-1}(\star\, w)\;,  
 \ee
where under the integral we can take ${\rm d}^{-1}(\star\, w)\equiv\omega_w$. The ambiguity due to $\omega_w$ being only 
well-defined up to a closed one-form is immaterial under the above integral thanks to $\star\, v$ being closed. 
Given this bilinear form, we can write a conventional (but non-local) Chern-Simons action for the Yang-Mills gauge fields $s_{\mu}$, 
 \be\label{convCS}
  S_{\rm CS} \ = \ \int {\rm d}^3x\,\varepsilon^{\mu\nu\rho}\big\langle s_{\mu},\partial_{\nu}s_{\rho}\ - \ \tfrac{1}{3}[s_{\nu},s_{\rho}]\big\rangle\;. 
 \ee
Noticing that the previously introduced gauge fields are according to (\ref{newnot}) related by 
 \be
  \star \,s_{\mu} \ = \ {\rm d}A_{\mu}\qquad \Leftrightarrow \qquad {\rm d}^{-1}(\star\,s_{\mu}) \ = \ A_{\mu}\;, 
 \ee
it is then easy to verify, using (\ref{dinversetheta}), (\ref{crossotoLIE}),  that this Chern-Simons action is equivalent to (\ref{finalBTCS}). 
Again, the advantage of the Leibniz-Chern-Simons formulation that treats the one-form-valued gauge fields as fundamental 
is that this formulation is manifestly local and does not require any topological assumptions
beyond the existence of a volume form. Thus, the formulation based on Leibniz algebras exists for any 
orientable 3-manifold. Moreover, using (\ref{invariantmetric}) one may formulate $\frak{sdiff}_3$ Yang-Mills theories in, say, $D=4$, 
and it would be interesting to investigate whether there are similar `dual' formulations that circumvent topological constraints.

After this digression, let us now turn to the matter couplings. The representation space is given by the space of functions (zero-forms) on 
the 3-manifold, $R={\cal C}^{\infty}(M)$, with the invariant metric defined for functions $f, g\in {\cal C}^{\infty}(M)$ by the integral 
 \be\label{sclaarmetric}
  \langle f, g\rangle \ \equiv \ \int {\rm d}^3y\,e\, f \, g\;. 
 \ee
In order to determine the map $\pi:R\wedge R\rightarrow \frak{g}^*$ defined in (\ref{BestPI}) 
we compute 
 \be
 \begin{split}
  \pi(f\wedge g)(v) \ &= \ \int {\rm d}^3y\, e\, \pi(f\wedge g)_i \,v^i \ =  \ \langle f, {\cal L}_v g\rangle  \\
  \ &= \ \int {\rm d}^3y\,e\, f v^i\partial_i g 
  \ = \ \frac{1}{2}\int {\rm d}^3y\,e\,\big(f\partial_i g-g\partial_if\big)\,v^i\;, 
 \end{split} 
 \ee
from which we read off 
 \be\label{PiDEF}
  \pi(f\wedge  g) \ = \ \tfrac{1}{2}(f\,{\rm d}g-g\,{\rm d}f) \ \in \ \frak{g}^*\;. 
 \ee
In particular, ${\rm d}\pi(f\wedge  g) = {\rm d}f\wedge {\rm d}g$.  
We can then determine the invariant 4-tensor, recalling $\Theta(\omega, \eta)=-\int \omega\wedge {\rm d}\eta$, as follows 
 \be
  \{f,g,h,k\} \ \equiv \ \Theta(\pi(f,g), \pi(h,k)) \ = \ - \int f\,{\rm d}g\wedge {\rm d}h\wedge {\rm d}k\;. 
 \ee
Next, we derive from this a 3-bracket, i.e., a map $R\otimes R\otimes R\rightarrow R$, 
by use of the invariant metric (\ref{sclaarmetric}) according to 
 \be
  \{f,g,h,k\} \ = \ \langle \{f,g,h\}, k\rangle 
  \ = \ \int {\rm d}^3y\,\varepsilon^{ijk}\,\partial_if\,\partial_jg\,\partial_kh\, k\;, 
 \ee
where we integrated by parts. We now read off the 3-bracket 
 \be\label{NAMBU}
  \{f,g,h\}  
  \ = \ \epsilon^{ijk}\,\partial_i f\,\partial_jg\,\partial_kh \;, 
 \ee
which is precisely the Nambu bracket \cite{Nambu:1973qe}.
With these expressions for the 3-bracket and inner product it is easy to verify that the 
${\cal N}=8$ Lagrangian (\ref{totalLagrangian}) precisely agrees with that given in \cite{Bandos:2008jv}.

\section{Consistent truncations and deformations} 

In this section we consider consistent truncations of SDiff$_3$ theories. In the first subsection 
we take the 3-manifold to be $S^3$ and discuss the expansion into a complete set of spherical harmonics, together 
with possible consistent truncations to a subset of modes. We show that the only consistent truncation to a finite subset 
of $S^3$ modes yields the BLG model with $SO(4)$ gauge group. In the second subsection we take the 3-manifold to be 
$S^2\times S^1$ and show that a Scherk-Schwarz reduction on $S^1$, which breaks conformal invariance, 
yields a Yang-Mills theory with gauge group SDiff$_2$, which is the $N\rightarrow \infty$ limit of $SU(N)$.

\subsection{Spherical harmonics on $S^3$ and the BLG model}

We consider the 3-manifold $S^3$, which can be embedded into $\mathbb{R}^4$ through four embedding coordinates $Z^a(y)$, $a=1,\ldots, 4$, 
 \be\label{square1}
  Z^aZ^a \ = \ 1\;. 
 \ee 
A simple explicit parameterization is given by 
 \be\label{Zapara}
  Z^a \ = \ \big(y^1, y^2, y^3, \sqrt{1-|y|^2}\big)\;, 
 \ee 
where $|y|$ denotes the euclidean 3-norm. The volume measure $e$ can then be obtained from the 
round sphere metric, which is the induced metric 
 \be\label{S3metric}
  g_{ij} \ \equiv \ \partial_iZ^a\partial_jZ^a \ = \ \delta_{ij} + \frac{y^iy^j}{1-|y|^2}\;, 
 \ee 
via 
 \be\label{volume}
  e \ \equiv \ \sqrt{\det g} \ = \ \frac{1}{\sqrt{1-|y|^2}}\;. 
 \ee 
It should be recalled that only this volume measure is needed in order to define the SDiff$_3$ theory, and 
the metric should therefore be viewed as an auxiliary object.  
The $Z^a$ satisfy further relations. First, they are orthonormal in that 
 \be\label{NORMMM}
  \langle Z^a, Z^b\rangle \ = \ \int {\rm d}^3y\,e\, Z^a Z^b \ = \ \frac{1}{4}\, \delta^{ab}\;.  
 \ee
Here a comment on normalization is in order. We re-interpret all integrals of the previous subsection to 
be normalized by the volume of the 3-sphere of unit radius, $\int\rightarrow \tfrac{1}{2\pi^2}\int$, 
including the integral defining the pairing  (\ref{covectorvector})  between vector and covector 
and that defining the inner product  (\ref{sclaarmetric}). The normalization of (\ref{NORMMM}) is then compatible with (\ref{square1}). 
Next, with (\ref{Zapara}) and (\ref{volume}) we have $Z^4=e^{-1}$, $\partial_iZ^4=-eZ^i$, $i=1,2,3$, and using these equations one may quickly 
verify that the Nambu bracket (\ref{NAMBU}) for the embedding coordinates takes the simple form 
  \be\label{NBBLG}
  \{ Z^a, Z^b, Z^c\} \ = \ \varepsilon^{abcd} Z^d\;, 
 \ee
in terms of the $SO(4)$ invariant epsilon tensor \cite{Bagger:2007vi,Axenides:2008rn}.

We next turn to the complete set of spherical harmonics on $S^3$, which can be directly constructed in terms of 
polynomials of arbitrary degrees in the $Z^a$. The scalars $\phi$ in the representation space ${\cal C}^{\infty}(M)$ can 
then be expanded as 
 \be\label{harmonics}
  \phi(x,y) \ = \ \sum_{\ell=0}^{\infty} \phi_{a_1\ldots a_{\ell}}(x) \,Z^{a_1}(y)\cdots Z^{a_{\ell}}(y)\;, 
 \ee 
where the coefficients $\phi_{a_1\ldots a_{\ell}}$ are totally symmetric and traceless, owing to the constraint (\ref{square1}). 
The expansion for the fermions $\psi$ is completely analogous. The expansion for the gauge fields reads 
 \be\label{vectorEXPAN}
   A_{\mu i}(x,y) \ = \ \sum_{\ell =1}^{\infty} A_{\mu\, a_{1}\ldots a_{\ell},b}(x)Z^{a_1}\cdots Z^{a_{\ell}}\partial_i Z^b\;, 
 \ee
where the coefficients live in the Young-tableaux representation $(\ell,1)$ under $SO(4)$, i.e., they are traceless and 
satisfy $A_{\mu (a_{1}\ldots a_{\ell},b)}=0$. The totally symmetric part is a total internal derivative and hence pure 
gauge w.r.t.~the gauge symmetry encoded in the last term in (\ref{gaugeAform}). 
Such expansions make it plain that, from the viewpoint of a 3D field theory, the SDiff$_3$ theory carries an infinite number 
of component fields with an infinite-dimensional gauge group. 

Let us next ask whether there are consistent truncations for which only a subset of the modes in (\ref{harmonics}) is included. 
A necessary consistency condition is that the Nambu bracket (\ref{NAMBU}) closes on this subset: the bracket of 
modes within the subset should again be in the subset. This is required, for instance, so that the supersymmetry rules (\ref{SUSYrules}) 
that employ the Nambu bracket yield consistent variations for the truncated fields. 
Of course, the simplest consistent truncation is to the zero-modes, i.e., to fields  independent of $y$. 
However, since the gauge structure is encoded in the internal derivatives, this truncation yields  the 
free theory (\ref{freeLagrangian}) and is hence of little interest. 
In order to obtain non-trivial truncations, we use the relation (\ref{NBBLG}) to 
infer that the Nambu bracket of polynomials of degrees $\ell_1$, $\ell_2$ and $\ell_3$ yields a polynomial of degree $\ell_1+\ell_2+\ell_3-2$. 
Thus, restricting to a finite set of modes of highest degree $\ell$ is only closed (and thereby consistent) for $\ell=1$. We will 
see below that this truncation yields the BLG model. 
On the other hand, if one allows for infinite truncations there are more possibilities as  
one may truncate, for instance, to polynomials of only even degree. We leave a more detailed investigation of such truncations 
and their physical significance for future work. 

We now return to the unique non-trivial finite truncation of (\ref{harmonics}) linear in the $Z^a$, 
for which the scalar and spinor fields are subjected to the (generalized Scherk-Schwarz) ansatz: 
 \be\label{scalarspinorAnsatz}
  \phi(x,y) \ = \ \phi^a(x) Z^a(y)\;, \qquad \psi(x,y) \ = \ \psi^a(x) Z^a(y)\;. 
 \ee
Here we could have included zero-mode terms, adding  $\varphi(x)$ to the scalar ansatz and similarly for the fermions, 
but it turns out that such fields do not couple to the $\phi^a$ or $\psi^a$ as, for instance, the Nambu bracket vanishes whenever one 
argument is independent of $y$. Thus, $\varphi$ does not contribute to the scalar potential, and it may be verified more generally 
that these fields decouple. Thus, it is sufficient to start with the ansatz (\ref{scalarspinorAnsatz}). 
For the vector fields, taking values in $\frak{g}^*$, the Scherk-Schwarz ansatz is given by truncating to the first term in  (\ref{vectorEXPAN}), which 
in terms of the map $\pi:R\wedge R\rightarrow \frak{g}^*$ 
defined in (\ref{PiDEF}) yields the matrix-valued one-form 
 \be\label{UMatrix}
  U^{ab}(y) \ = \ \pi(Z^a(y), Z^b(y)) \ = \ Z^{[a}\partial_i Z^{b]} {\rm d}y^i\;, 
 \ee
so that we can write for the gauge field  
 \be\label{vectoransatz}
  A_{\mu i}(x,y) \ = \ A_{\mu ab}(x) U_i{}^{ab}(y)\;. 
 \ee
 
In the following we will see that the action and supersymmetry rules reduce consistently
as a consequence of a consistency condition that can be viewed as 
generalized parallelizability \cite{Lee:2014mla,Baguet:2015sma} expressed in terms of  the Leibniz algebra structure on $\frak{g}^*$. 
Specifically, in terms of the Leibniz product (\ref{symmRELLL2}) we require
 \be\label{genPar}
  U_{ab}  \circ  U_{cd} \ = \ -X_{ab,cd}{}^{ef} U_{ef}\;,
  \ee
where the $X$ are structure constants,  
\be
X_{ab,cd}{}^{ef}\ \equiv \ \Theta_{ab,gh}f^{gh, ef}{}_{cd}\;, 
\ee 
expressed in terms of the embedding tensor  for $\frak{g}=\frak{so}(n)$, whose  
structure constants are $f^{ab,cd}{}_{ef}=-2\delta^{[a}{}_{[e}\delta^{b][c}\delta^{d]}{}_{f]}$. 
Only for $n=4$ can the quadratic and linear constraints for the embedding tensor be solved, in which case 
$\Theta_{ab,cd}=\varepsilon_{abcd}$, and hence 
 \be\label{altSO(4)}
  X_{ab,cd}{}^{ef} \ = \ -2\, \varepsilon_{ab[c}{}^{[e}\,\delta^{f]}{}_{d]}\;. 
 \ee
We next prove that for $U^{ab}$ defined in (\ref{UMatrix}) the generalized parallelizability (\ref{genPar}) indeed holds  
for these structure constants.  We first note that the one-forms $U^{ab}$ are mapped under $\vartheta$ to the vector fields 
 \be
  \vartheta(U^{ab})^i \ = \ \epsilon^{ijk}\partial_j(Z^{[a}\partial_k Z^{b]}) \ = \ \epsilon^{ijk}\partial_j Z^a \partial_k Z^b\;. 
 \ee
Thus, acting via the Lie derivative on another embedding scalar yields 
 \be\label{UactingonZ}
  {\cal L}_{\vartheta(U^{ab})}Z^c \ = \ \vartheta(U^{ab})^i \partial_i Z^c \ = \ \{Z^a, Z^b, Z^c\} \ = \ \varepsilon^{abcd} Z^d\;, 
 \ee
where we used (\ref{NBBLG}).  
Since the definition (\ref{UMatrix}) of $U^{ab}$ is manifestly diffeomorphism covariant, we can immediately compute 
the Leibniz product from the 
defintion (\ref{oneThetaDEF}),  
 \be
 \begin{split}
  U^{ab}\circ U^{cd} \ &= \ {\cal L}_{\vartheta(U^{ab})} U^{cd} \ = \ ({\cal L}_{\vartheta(U^{ab})}Z^{[c})\,{\rm d} Z^{d]}
 + Z^{[c}\,{\rm d} ({\cal L}_{\vartheta(U^{ab})}Z^{d]}) \\
 \ &= \ \varepsilon^{ab[c}{}_{e} Z^{|e|}\,{\rm d} Z^{d]}  + \varepsilon^{ab[d}{}_{e} Z^{c]}\,{\rm d}Z^e \\
 \ &= \ 2\,\varepsilon^{ab[c}{}_{e} \,U^{|e|d]}\;. 
 \end{split}
 \ee
This proves (\ref{genPar}) for (\ref{altSO(4)}). 
We can also determine the symmetric bilinear form of the embedding tensor, 
obtained by evaluating it on the $U_{ab}$, to confirm $\Theta(U_{ab}, U_{cd}) \propto \varepsilon_{abcd}$. 
To this end we compute from the definition (\ref{embeddingAGAIN}) 
 \be\label{Thetaabcd}
 \begin{split}
  \Theta(U^{ab}, U^{cd}) \ &= \ -\int {\rm d}^3y\,\varepsilon^{ijk} U_{i}{}^{ab}\partial_j U_{k}{}^{cd}
  \ = \ -\int {\rm d}^3y\,\varepsilon^{ijk} Z^{[a}\partial_i Z^{b]}\partial_j Z^c \partial_k Z^d \\
  \ &= \ -\int {\rm d}^3y\,e\,Z^{[a} \{Z^{b]}, Z^c, Z^d\} \ = \ \tfrac{1}{4}\,\varepsilon^{abcd}\;, 
 \end{split}
 \ee
using (\ref{NORMMM}) in the last step.

We will now argue that the Bandos-Townsend theory for $S^3$ consistently reduces 
to the BLG model. Abstractly, this follows immediately from the observation in  \cite{Bergshoeff:2008cz} that the BLG model 
fits into the embedding tensor approach, with precisely the embedding tensor $\Theta_{ab,cd}\propto \varepsilon_{abcd}$ 
that was produced by the above reduction ansatz. However, for the convenience of the reader we illustrate in the following 
the consistent reduction of various structures.\footnote{See also \cite{Arvanitakis:2015sgs} for a proof using harmonic analysis that the 
Chern-Simons-like action of the Bandos-Townsend theory reduces to that of the BLG model.}

 First, thanks to the relation (\ref{NBBLG}) it is clear that all terms involving the 3-bracket 
(or alternatively the 4-tensor) reduce consistently. For the covariant derivatives on, say, the scalars one quickly finds 
with (\ref{vectoransatz}) and (\ref{UactingonZ}) 
 \be
  \begin{split}
   D_{\mu}\phi \ &= \ \partial_{\mu}\phi-\vartheta(A_{\mu})^i\partial_i\phi \\
   \ &= \ \partial_{\mu}\phi^a\,Z^a \ - \ A_{\mu ab}\{Z^a,Z^b,Z^c\}\phi^c\\
   \ &= \ D_{\mu}\phi^a Z^a\;, 
  \end{split}
 \ee  
where 
 \be
  D_{\mu}\phi^a \ = \ \partial_{\mu}\phi^a \ + \ A_{\mu \,cd}\,\varepsilon^{abcd} \phi_b\;, 
 \ee
which is the expected covariant derivative of the BLG model. Thus, the covariantized kinetic terms reduce correctly. 
Next, let us verify that the gauge field and its supersymmetry transformation reduces consistently. 
For the latter one computes with (\ref{SUSYrules}) and (\ref{vectoransatz}) 
 \be
 \begin{split}
  \delta_{\epsilon}A_{\mu i}(x,y) \ &= \ -i\bar{\epsilon}\,\gamma_{\mu}\,\bar{\Gamma}^I\pi(\psi^aZ^a,\phi^{Ib}X^b)_i\\
  \ &= \ -i\bar{\epsilon}\,\gamma_{\mu}\,\bar{\Gamma}^I\psi^a \phi^{Ib} \,U_{i}{}^{ab}\\
  \ &\equiv \ \delta_{\epsilon}A_{\mu\, ab}(x)\, U_{i}{}^{ab}(y)\;, 
 \end{split} 
 \ee
from which we infer that the supersymmetry variation reduces consistently, with 
 \be
  \delta_{\epsilon}A_{\mu\, ab} \ = \ -i\bar{\epsilon}\,\gamma_{\mu}\,\bar{\Gamma}^I\psi_{[a}\, \phi_{b]}^I\;, 
 \ee
in agreement with the expected supersymmetry rule for the gauge fields of the BLG model.  
For the Chern-Simons term we note that, thanks to the $U^{ab}$ satisfying the algebra (\ref{genPar}),  
the Leibniz-Chern-Simons action (\ref{secondorthirdCS}) reduces to the Chern-Simons term of 
the BLG model with bilinear embedding tensor (\ref{Thetaabcd}). 
This completes our proof that the BLG model is a consistent truncation of the Bandos-Townsend theory based on 
the 3-manifold $S^3$ and, moreover, we determined the uplift formulas to be  given by (\ref{scalarspinorAnsatz}) and (\ref{vectoransatz}).

We close this subsection by discussing the algebraic structure encoded in (\ref{genPar}) in a little more detail. 
In particular, we will clarify that, in contrast to the full SDiff$_3$ theory and despite appearance,  the BLG model is 
not a `higher gauge theory' but rather based on a conventional Lie algebra. 
Let us first point out that the one-forms $U_{i}{}^{ab}$ defined in (\ref{UMatrix}) are naturally related to the 
Killing vector fields of the round $S^3$ metric, which can be written as 
 \be
  K^{i \,ab} \ \equiv \ g^{ij} U_{j}{}^{ab}\;, 
 \ee
using the metric (\ref{S3metric}) to map a one-form to a vector. Being Killing vector fields of $S^3$ they naturally satisfy 
the $\frak{so}(4)$ algebra w.r.t.~the Lie bracket,  
 \be\label{Killingalgbera}
  [K^{ab}, K^{cd}] \ = \ f^{ab,cd}{}_{ef} K^{ef}\;. 
 \ee
These Killing vector fields enter the ansatz for Kaluza-Klein vector fields for compactifications of (super-)gravity on $S^3$, 
which thanks to this algebra relation yields an $SO(4)$ gauge symmetry. The relation (\ref{genPar}) is the analogue 
to (\ref{Killingalgbera}) that, however, differs in two important (and related) respects: \textit{i)} the standard $\frak{so}(4)$ structure constants 
enter (\ref{Killingalgbera}), while the `epsilon-twisted' structure constants (\ref{altSO(4)}) enter (\ref{genPar}); 
\textit{ii)} while the algebraic operation on the left-hand side of (\ref{Killingalgbera}) is the Lie bracket of vector fields, 
the structure on the left-hand side of (\ref{genPar}) is the novel Leibniz product on one-forms. 

Despite these differences, and despite the fact that a Leibniz algebra generically does not define  a Lie algebra, 
we will now show that the `epsilon-twisted' structure constants (\ref{altSO(4)}) emerging from the truncation 
define a genuine Lie algebra. 
First, we write out  the Leibniz product as follows 
 \be
  (V\circ W)^{ab} \ \equiv \ V^{cd} W^{ef} X_{cd, ef}{}^{ab} \ = \ 
  2\,V^{cd} \,W^{e[a}\,\varepsilon_{cde}{}^{\,b]}\;. 
 \ee
This product, which by construction defines a Leibniz algebra, is not manifestly antisymmetric. We will now show, however, 
that it is secretly antisymmetric and hence actually defines a genuine Lie algebra. 
To this end note that by the polarization relation 
 \be
  \{V,W\} \ = \ \frac{1}{2}\big( (V+W)\circ (V+W)-V\circ V-W\circ W\big)\;, 
 \ee 
the symmetrized product vanishes if and only 
if the product vanishes on diagonal arguments. Denoting such an element by $U^{ab}$ it follows  from the Schouten identity $U_{e}{}^{[a}\,\varepsilon^{bcde]}\equiv 0$ together 
with $U^e{}_{e}=0$ that 
 \be
  (U\circ U)^{ab} \ = \ -2\, U_{cd} \,U_{e}{}^{[a} \,\varepsilon^{b]cde} \ = \ 2\, U_{cd} \, U_{e}{}^{[c}\,\varepsilon^{d]abe}
  \ = \  2\, (U_{cd} \, U_{e}{}^{c})\,\varepsilon^{dabe} \ = \ 0\;, 
 \ee
using in the last step that the expression in parenthesis is manifestly symmetric in its two external indices $d, e$. 
This shows that the gauge algebra of BLG is governed by a genuine Lie algebra. In particular, 
the 3-bracket that encodes the gauge structure in the sense of L$_{\infty}$ algebras 
\textit{vanishes identically for the BLG model}. Of course, this bracket should not be confused with the `3-bracket' 
on scalar fields, which defines in particular the scalar potential but has nothing directly to do with the gauge algebra.

The fact that there are two different sets of structure constants characterizing the same gauge group $SO(4)$
can be understood as a consequence of  the Lie algebra $\frak{g}=\frak{so}(4)=\frak{su}(2)\oplus \frak{su}(2)$ 
not being semi-simple and having two isomorphisms between $\frak{g}$ and $\frak{g}^*$. The first, $\frak{i}:\frak{so}(n)^*\rightarrow \frak{so}(n)$,  
exists for any $SO(n)$ groups and is given by $\frak{i}(\tilde{t}_{ab})  =  \delta_{c[a}\delta_{b]d} \,t^{cd}$, where $t^{ab}$ 
are the generators and $\tilde{t}_{ab}$ are the dual generators. 
For $\frak{g}=\frak{so}(4)$, however, there is a second isomorphism, $\vartheta :\frak{g}^*\rightarrow \frak{g}$, given by the embedding tensor,  
 \be 
   \vartheta(\tilde{t}_{ab}) \ = \ \tfrac{1}{2}\,\varepsilon_{abcd}\,t^{cd}\;. 
 \ee
Thanks to $\vartheta$ being an isomorphism, the Lie algebra structure on $\frak{so}(4)$ can be transported to 
a Lie algebra structure on $\frak{so}(4)^*$, which defines the gauge algebra of the BLG model.

\subsection{Scherk-Schwarz reduction and SDiff$_2$ Yang-Mills theory}

The above truncation preserves the conformal symmetry and the $SO(8)$ R-symmetry. We now consider a 
Scherk-Schwarz compactification that breaks the conformal symmetry and breaks the R-symmetry down to $SO(7)$, 
which will lead to a 3D super-Yang-Mills theory. This example was already discussed in \cite{Bandos:2008jv}, but for completeness 
we now review, in the present formulation,  the essential features of this analysis. 
We temporarily relabel the $SO(8)$ index as $\hat{I}=(I,8)$, where now 
$I=1,\ldots, 7$ is an $SO(7)$ index.  
Moreover, we take the 3-manifold to be $M_2\times S^1$, so that one can choose global coordinates 
 \be\label{coorddecompose}
  y^i \ = \ (\sigma^{\alpha}, z)\;, \quad \alpha=1,2\;, 
 \ee
where $\sigma^{\alpha}$ are coordinates on $M_2$. Natural choices for $M_2$ are $T^2$ or $S^2$, but the following 
discussion does not depend on the topology of the 2-manifold.   
The Scherk-Schwarz ansatz singles out the eighths scalar 
and takes the form
 \be\label{ScalarSSansatz}
  \phi^8(x,y) \ = \ \varphi(x,\sigma) + \sqrt{m} z\;, 
 \ee 
where $m$ is a mass scale, in agreement with the mass dimensions of the scalars being $\frac{1}{2}$.  
Thus, this ansatz breaks conformal symmetry, and it breaks $SO(8)$ to $SO(7)$. 
The Scherk-Schwarz anzatz for all other scalar and spinor fields is trivial, declaring  them to be independent of $z$.

Next, we turn to the reduction of the gauge vectors, which is a little more subtle in that only the `derived' $\frak{sdiff}_3$ gauge fields 
$s_{\mu}\equiv \vartheta(A_{\mu})$ have a local reduction formula, which simply states that the $s_{\mu}$ are independent of $z$, 
 \be\label{vectoransatzzz}
  s_{\mu}{}^{i}(x,y) \ = \ (s_{\mu}{}^{\alpha}(x,\sigma), s_{\mu}{}^{z}(x,\sigma)) \ \equiv \ 
  (\epsilon^{\alpha\beta}\partial_{\beta} a_{\mu}(x,\sigma), b_{\mu}(x,\sigma))\;. 
 \ee
For completeness we briefly discuss the non-local ansatz for the fundamental fields. 
To this end we note that  $s_{\mu}{}^{i}=\epsilon^{ijk}\partial_j A_{\mu k}$ decomposes under (\ref{coorddecompose}) as 
 \be\label{sSTEPPP}
  s_{\mu}{}^{i} \ = \ (\epsilon^{\alpha\beta}\partial_{\beta} A_{\mu z} - \epsilon^{\alpha\beta} \partial_z A_{\mu\beta}\,,\; 
  \epsilon^{\alpha\beta}\partial_{\alpha} A_{\mu\beta})\;, 
 \ee
so that by comparison with (\ref{vectoransatzzz}) 
 \be
  b_{\mu} \ = \ \epsilon^{\alpha\beta}\partial_{\alpha} A_{\mu\beta}\;. 
 \ee
This can be solved, non-locally, to give the ansatz for the $A_{\mu}$: 
 \be
 \begin{split}
  A_{\mu \alpha}(x,y) \ &= \ - \tfrac{1}{2}\,\epsilon_{\alpha\beta}\int^{\sigma} {\rm d}\sigma'^{\beta}\, b_{\mu}(x,\sigma')\;, \\
  A_{\mu z} (x,y) \ &= \ a_{\mu}(x,\sigma)\;, 
 \end{split} 
 \ee
assuming appropriate boundary conditions. We used in the second line that the first does not depend on $z$, which with (\ref{sSTEPPP}) 
gives the ansatz for $A_{\mu z}$. 
In the following we will see that the gauge symmetry reduces to SDiff$_2$, for which $a_{\mu}$ will act as the gauge field. 

Having specified the reduction ans\"atze, let us now work out how the various terms in the action reduce. 
First, the covariant derivatives of the first seven scalars reduce as 
 \be
  D_{\mu}\phi^I \ = \ \partial_{\mu}\phi^I -\{a_{\mu},\phi^I\}\;, 
 \ee
where 
 \be
 \{A, B\} \ = \ -\,\epsilon^{\alpha\beta}\,\partial_{\alpha} A\,\partial_{\beta} B
 \ee 
are Poisson brackets defining the Lie algebra $\frak{sdiff}_2$.
Thus, one obtains the covariant derivative w.r.t.~$\frak{sdiff}_2$.
Next, for the eighths scalar $\phi^8$ the covariant derivative reduces as 
 \be
 \begin{split}
  D_{\mu}\phi^8 \ &= \ \partial_{\mu}\phi^8 - s_{\mu}{}^{\alpha}\partial_{\alpha}\phi^8 - s_{\mu}{}^{z}\partial_z\phi^8 \;, 
 \end{split} 
 \ee
from which we compute with (\ref{ScalarSSansatz}) 
 \be\label{shiftcovariant}
  D_{\mu}\varphi \ = \ \partial_{\mu}\varphi - \{a_{\mu}, \varphi\} - \sqrt{m}\, b_{\mu}\;. 
 \ee
From this we infer that there is a  St\"uckelberg gauge symmetry, for which the $b_{\mu}$ are the gauge fields, 
 \be
  \delta_\chi \varphi \ = \ \sqrt{m} \chi\,, \quad  \delta_{\chi}b_{\mu} \ = \ D_{\mu}\chi\;. 
 \ee 
Thus,  $\varphi$ is a `Goldstone boson' that can be gauged away, which in turn 
generates a Yang-Mills kinetic term, as we will discuss now.  

To this end, we observe that the Chern-Simons term reduces as 
 \be
  \begin{split}
   S \ &= \ -\int {\rm d}^3x\int {\rm d}^3y\,\varepsilon^{\mu\nu\rho} \big(A_{\mu i}\partial_{\nu}s_{\rho}{}^{i} 
   +\tfrac{1}{3}\epsilon_{ijk} s_{\mu}{}^{i} s_{\nu}{}^{j} s_{\rho}{}^{k}\big)
   \\
     \ &= \ -\int {\rm d}^3x\int {\rm d}^3y\,\varepsilon^{\mu\nu\rho}\,b_{\mu}f_{\nu\rho}(a)\;, 
  \end{split}
 \ee  
where 
 \be\label{sdiff2fieldstrength}
  f_{\mu\nu} \ = \ \partial_{\mu}a_{\nu}-\partial_{\nu}a_{\mu} - \{a_{\mu},a_{\nu}\}
 \ee
is the Yang-Mills field strength for $\frak{sdiff}_2$. Thus, we obtain a BF-type coupling. 
It is easy to see that in an action consisting of a kinetic term based on (\ref{shiftcovariant}) and a BF-type 
Chern-Simons term one may eliminate  $b_{\mu}$ algebraically, which generates a Yang-Mills term 
for (\ref{sdiff2fieldstrength}) \cite{Nicolai:2003bp}. 

For the remaining couplings in the action and the supersymmetry variations we have to inspect 
the reduction of the 3-bracket, which yields the $\frak{sdiff}_2$ Lie bracket: 
 \be
  \{\phi^I, \phi^J, \phi^8\} \ = \ -\sqrt{m}\{\phi^I, \phi^J\}\;. 
 \ee
It then follows that the potential and Yukawa-couplings give the expected contribution for a super-Yang-Mills theory 
with gauge algebra $\frak{sdiff}_2$. In order to verify the precise match also for the fermions it remains  
to decompose the  $SO(8)$ spinor representation under the surviving R-symmetry $SO(7)$. 
Luckily, these have the same dimensions, and 
so the spinors need not be decomposed. We refer to  \cite{Bandos:2008jv} for more details. 

Summarizing, the above Scherk-Schwarz ansatz provides a consistent truncation 
of the Bandos-Townsend theory for $M_2\times S^1$, which in turn is exactly equivalent to the 3D ${\cal N}=8$
super-Yang-Mills theory with gauge group SDiff$_2$. Let us emphasize that this relation is somewhat different 
from that discussed in \cite{Mukhi:2008ux} for the BLG model, where the analogue of the vector field $b_{\mu}$
also enters with cubic couplings, so that it cannot be integrated out exactly, but at best in some limit.

\subsection{Massive deformations} 

We now turn to a massive deformation of the general ${\cal N}=8$ SCFTs given in sec.~2, which again can be 
formulated using only the general structures given there. In particular, this yields a massive deformation of the 
Bandos-Townsend theory. In this we closely follow the massive deformation constructed in \cite{Gomis:2008cv,Hosomichi:2008qk}
for the BLG model.

The Lagrangian (\ref{totalLagrangian}) is deformed by the following terms, which are proportional to a mass parameter $m$, 
 \be\label{massLagrangian}
 \begin{split}
  {\cal L}_{m} \ = \ &-\tfrac{1}{2}m^2\langle \phi^I,\phi^I\rangle +\tfrac{i}{2}m\langle \bar{\psi},\Gamma_{1234}\psi\rangle 
  -\tfrac{1}{48}\, m\,
{\rm tr}\left[ \Gamma^{1234}\Gamma^{IJKL} \right]
\left\{\phi^{{I}},\phi^{{J}},\phi^{{K}}, \phi^{{L}}\right\}\;, 
 \end{split} 
 \ee
where $\Gamma^{1234}=\Gamma^{1}\bar{\Gamma}^2\Gamma^{3}\bar{\Gamma}^4$,  
such that the
R-symmetry group is broken to $SO(4)\times SO(4)$. The supersymmetry transformations (\ref{SUSYrules}) only get deformed on the spinor fields, whose 
variations receive the following additional term: 
 \be
  \delta_{\epsilon}'\psi \ = \ m\Gamma_{1234}\Gamma^I\phi^I\epsilon\;. 
 \ee
 
In the following we discuss the proof of ${\cal N}=8$ supersymmetry.  
The variations ${\cal O}(m^2)$ of the scalar and fermion mass terms cancel directly due to $(\Gamma_{1234})^2=1$.  The variations ${\cal O}(m)$ and linear in $\phi$
originate from the $\delta_{\epsilon}$ variation of the fermionic mass term and the $\delta_{\epsilon}^{\prime}$ variation of the fermion kinetic term, 
which cancel directly. 
The most subtle variations are of order  ${\cal O}(m)$ and cubic in $\phi$. They originate from: \\
\textit{i)} the higher-order variation of the fermion mass term, 
 \be
  \begin{split}
  \delta_{\epsilon}{\cal L}_{\psi,{\rm mass}} \ = \ & \tfrac{1}{6}\,i m \{\phi^I,\phi^J,\phi^K,\bar{\psi}\}\Gamma^{1234}\Gamma^{IJK}\epsilon
  \\
   \ = \ &  \tfrac{1}{32}\,i m \,{\rm tr}\left[ \Gamma^{1234}\Gamma^{IJKL} \right]
   \left(
   \tfrac{2}{3}
      \left\{\phi^I,\phi^J,\phi^K,\bar{\psi}\right\}\Gamma^{L}
   -\left\{\phi^I,\phi^J,\phi^P,\bar{\psi}\right\}\Gamma^{PKL}  
   \right) \epsilon
   \;,
    \end{split}
 \ee
\textit{ii)} the $\delta_{\epsilon}'$ variation of the Yukawa couplings: 
 \be
  \begin{split}
  \delta'_{\epsilon}{\cal L}_{\rm Yukawa} \ = \ & \tfrac{1}{2}\,im \{\phi^I,\phi^J,\phi^K,\bar{\psi}\}\Gamma^{IJ}\Gamma^{1234}\Gamma^{K}\epsilon
    \\
   \ = \ &  \tfrac{1}{32}\,i m \,{\rm tr}\left[ \Gamma^{1234}\Gamma^{IJKL} \right]
   \left(
   2
      \left\{\phi^I,\phi^J,\phi^K,\bar{\psi}\right\}\Gamma^{L}
   +\left\{\phi^I,\phi^J,\phi^P,\bar{\psi}\right\}\Gamma^{PKL}  
   \right) \epsilon
   \;,
  \end{split}
 \ee
\textit{iii)} the $\delta_{\epsilon}$ variations of the quartic terms in (\ref{massLagrangian}): 
 \be
   \delta_{\epsilon}{\cal L}_{\rm quartic} \ = \ 
 -   \tfrac{1}{12}\,i m \,{\rm tr}\left[ \Gamma^{1234}\Gamma^{IJKL} \right]
      \left\{\phi^I,\phi^J,\phi^K,\bar{\psi}\right\}\Gamma^{L}\,\epsilon
\;,
 \ee
and precisely cancel.

Note that this massive deformation preserves ${\cal N}=8$ supersymmetry, i.e., 16 real supercharges. 
Nevertheless, the theory features only massive spin-0 and spin-$\frac{1}{2}$ excitations,  
which is possible thanks to a novel kind of multiplet shortening in 3D due to non-central charges in 
the Poincar\'e superalgebra \cite{Bergshoeff:2008ta}.

\section{Conclusions and Outlook} 
We have provided a general formulation of ${\cal N}=8$ SCFTs that includes 
infinite-dimensional gauge algebras, thereby circumventing the no-go results of \cite{Papadopoulos:2008sk,Gauntlett:2008uf}. 
Intriguingly, we find that not only do the BLG model and the Bandos-Townsend theories fit into the same general 
framework, the former is actually a (consistent) subsector of the latter. The Bandos-Townsend theories thus encode, for different choices of the 
3-manifold, the BLG model for $S^3$ on the one hand and the SDiff$_2$ super-Yang-Mills theory for $S^2\times S^1$
on the other. In addition,  one automatically obtains a generalization of the 
BLG model, whose gauge group $SO(4)$ is extended to an infinite-dimensional gauge symmetry. 
It is important to investigate whether there are other consistent truncations of theses theories, perhaps to models with less supersymmetry. 
Furthermore, it would be interesting to directly construct theories with less supersymmetry, which may still employ 
Leibniz-Chern-Simons actions. 

Most importantly, it remains to classify the solutions of the linear and quadratic constraints summarized in the introduction 
and thereby to classify the Lagrangian ${\cal N}=8$ superconformal field theories that fit into this framework. 
While so far we have not been able to construct a solution that is more general than the one leading to the 
Bandos-Townsend theories, a promising avenue should be to explore `fuzzy' spaces like the fuzzy 3-spheres that 
play a role in the proposal by Basu and Harvey \cite{Basu:2004ed}.

\subsection*{Acknowledgements}
We would like to thank Dennis Sullivan, Paul Townsend, Alan Weinstein and Barton Zwiebach for useful discussions. 

The work of O.H. is supported by an ERC Consolidator Grant.


\begin{thebibliography}{99}


\bibitem{Aharony:2008ug}
  O.~Aharony, O.~Bergman, D.~L.~Jafferis and J.~Maldacena,
  ``N=6 superconformal Chern-Simons-matter theories, M2-branes and their gravity duals,''
  JHEP {\bf 0810} (2008) 091
  doi:10.1088/1126-6708/2008/10/091
  [arXiv:0806.1218 [hep-th]].

\bibitem{Bagger:2007jr}
  J.~Bagger and N.~Lambert,
  ``Gauge symmetry and supersymmetry of multiple M2-branes,''
  Phys.\ Rev.\ D {\bf 77} (2008) 065008
  doi:10.1103/PhysRevD.77.065008
  [arXiv:0711.0955 [hep-th]].

\bibitem{Gustavsson:2007vu}
  A.~Gustavsson,
  ``Algebraic structures on parallel M2-branes,''
  Nucl.\ Phys.\ B {\bf 811} (2009) 66
  doi:10.1016/j.nuclphysb.2008.11.014
  [arXiv:0709.1260 [hep-th]].

\bibitem{Papadopoulos:2008sk}
  G.~Papadopoulos,
  ``M2-branes, 3-Lie Algebras and Plucker relations,''
  JHEP {\bf 0805} (2008) 054
  doi:10.1088/1126-6708/2008/05/054
  [arXiv:0804.2662 [hep-th]].
  
\bibitem{Gauntlett:2008uf} 
  J.~P.~Gauntlett and J.~B.~Gutowski,
  ``Constraining Maximally Supersymmetric Membrane Actions,''
  JHEP {\bf 0806}, 053 (2008)
  doi:10.1088/1126-6708/2008/06/053
  [arXiv:0804.3078 [hep-th]].
 



\bibitem{Bandos:2008jv} 
  I.~A.~Bandos and P.~K.~Townsend,
  ``SDiff Gauge Theory and the M2 Condensate,''
  JHEP {\bf 0902}, 013 (2009)
  doi:10.1088/1126-6708/2009/02/013
  [arXiv:0808.1583 [hep-th]].
  
  
\bibitem{Floratos:1988mh} 
  E.~G.~Floratos, J.~Iliopoulos and G.~Tiktopoulos,
  ``A note on SU(infinity) classical Yang-Mills theories,''
  Phys.\ Lett.\ B {\bf 217}, 285 (1989).
  doi:10.1016/0370-2693(89)90867-8


\bibitem{Hohm:2013pua} 
  O.~Hohm and H.~Samtleben,
  ``Exceptional Form of D=11 Supergravity,''
  Phys.\ Rev.\ Lett.\  {\bf 111}, 231601 (2013)
  doi:10.1103/PhysRevLett.111.231601
  [arXiv:1308.1673 [hep-th]].

\bibitem{Hohm:2013vpa} 
  O.~Hohm and H.~Samtleben,
  ``Exceptional Field Theory I: $E_{6(6)}$ covariant Form of M-Theory and Type IIB,''
  Phys.\ Rev.\ D {\bf 89}, no. 6, 066016 (2014)
  doi:10.1103/PhysRevD.89.066016
  [arXiv:1312.0614 [hep-th]].


\bibitem{Hohm:2013uia}
  O.~Hohm and H.~Samtleben,
  ``Exceptional field theory. II. E$_{7(7)}$,''
  Phys.\ Rev.\ D {\bf 89} (2014) 066017
  doi:10.1103/PhysRevD.89.066017
  [arXiv:1312.4542 [hep-th]].

\bibitem{Hohm:2014fxa} 
  O.~Hohm and H.~Samtleben,
  ``Exceptional field theory. III. E$_{8(8)}$,''
  Phys.\ Rev.\ D {\bf 90}, 066002 (2014)
  doi:10.1103/PhysRevD.90.066002
  [arXiv:1406.3348 [hep-th]].

\bibitem{Siegel:1993th}
  W.~Siegel,
  ``Superspace duality in low-energy superstrings,''
  Phys.\ Rev.\ D {\bf 48} (1993) 2826
  doi:10.1103/PhysRevD.48.2826
  [hep-th/9305073].
  
\bibitem{Hull:2009mi} 
  C.~Hull and B.~Zwiebach,
  ``Double Field Theory,''
  JHEP {\bf 0909}, 099 (2009)
  doi:10.1088/1126-6708/2009/09/099
  [arXiv:0904.4664 [hep-th]].  

\bibitem{Hull:2009zb}
  C.~Hull and B.~Zwiebach,
  ``The Gauge algebra of double field theory and Courant brackets,''
  JHEP {\bf 0909} (2009) 090
  doi:10.1088/1126-6708/2009/09/090
  [arXiv:0908.1792 [hep-th]].
  
\bibitem{Hohm:2010jy}
  O.~Hohm, C.~Hull and B.~Zwiebach,
  ``Background independent action for double field theory,''
  JHEP {\bf 1007} (2010) 016
  doi:10.1007/JHEP07(2010)016
  [arXiv:1003.5027 [hep-th]].  
  
\bibitem{Hohm:2010pp}
  O.~Hohm, C.~Hull and B.~Zwiebach,
  ``Generalized metric formulation of double field theory,''
  JHEP {\bf 1008} (2010) 008
  doi:10.1007/JHEP08(2010)008
  [arXiv:1006.4823 [hep-th]]. 
  
  
       \bibitem{TCourant}
  T.~Courant,
  ``Dirac Manifolds,"
   Transactions of the American Mathematical Society, 
   vol.~319, number 2, 1990. 
   
   
\bibitem{LieXu}
Z.~Liu, A.~Weinstein, P.~Xu, 
``Manin triples for Lie bialgebroids,"
J.~Diff.~Geom., 1995, 
[arxiv:dg-ga/9508013]. 
   

  
     \bibitem{roytenberg-weinstein}
  D.~Roytenberg and A.~Weinstein,
  ``Courant Algebroids and Strongly Homotopy Lie Algebras,"
   Lett. Math. Phys. {\bf 46} (1998), 81--93
 [arXiv:math/9802118]. 

 
 
      \bibitem{roytenberg}
  D.~Roytenberg,
  ``Courant Algebroids, derived brackets and even symplectic supermanifolds,"
 [arXiv:math/9910078]. 
 
 
 
\bibitem{Hohm:2018ybo} 
  O.~Hohm and H.~Samtleben,
  ``Leibniz-Chern-Simons Theory and Phases of Exceptional Field Theory,''
  arXiv:1805.03220 [hep-th].


\bibitem{Hohm:2017wtr} 
  O.~Hohm, E.~T.~Musaev and H.~Samtleben,
  ``O($d+1, d+1$) enhanced double field theory,''
  JHEP {\bf 1710}, no. 10, 086 (2017)
  doi:10.1007/JHEP10(2017)086
  [arXiv:1707.06693 [hep-th]].  
  
 \bibitem{LODAY}
J.-L. Loday, {\em Cyclic homology}, vol.~301 of {\em Grundlehren der
  Mathematischen Wissenschaften [Fundamental Principles of Mathematical
  Sciences]}.
\newblock Springer-Verlag, Berlin, 1992. 

\bibitem{Zwiebach:1992ie}
  B.~Zwiebach,
  ``Closed string field theory: Quantum action and the B-V master equation,''
  Nucl.\ Phys.\ B {\bf 390} (1993) 33
  doi:10.1016/0550-3213(93)90388-6
  [hep-th/9206084].


\bibitem{Hohm:2017pnh} 
  O.~Hohm and B.~Zwiebach,
  ``$L_{\infty}$ Algebras and Field Theory,''
  Fortsch.\ Phys.\  {\bf 65}, no. 3-4, 1700014 (2017)
  doi:10.1002/prop.201700014
  [arXiv:1701.08824 [hep-th]].
  
 \bibitem{Nicolai:2000sc} 
  H.~Nicolai and H.~Samtleben,
  ``Maximal gauged supergravity in three-dimensions,''
  Phys.\ Rev.\ Lett.\  {\bf 86}, 1686 (2001)
  doi:10.1103/PhysRevLett.86.1686
  [hep-th/0010076].

\bibitem{deWit:2002vt} 
  B.~de Wit, H.~Samtleben and M.~Trigiante,
  ``On Lagrangians and gaugings of maximal supergravities,''
  Nucl.\ Phys.\ B {\bf 655}, 93 (2003)
  doi:10.1016/S0550-3213(03)00059-2
  [hep-th/0212239].

\bibitem{deWit:2003ja} 
  B.~de Wit, I.~Herger and H.~Samtleben,
  ``Gauged locally supersymmetric D = 3 nonlinear sigma models,''
  Nucl.\ Phys.\ B {\bf 671}, 175 (2003)
  doi:10.1016/j.nuclphysb.2003.08.022
  [hep-th/0307006].
  
 \bibitem{Strobl} 
Thomas Strobl, ``Mathematics around Lie 2-algebroids and the tensor hierarchy in gauged supergravity",
talk at ``Higher Lie theory", University of Luxembourg, 2013.  
 
\bibitem{Bergshoeff:2008cz}
  E.~A.~Bergshoeff, M.~de Roo and O.~Hohm,
  ``Multiple M2-branes and the Embedding Tensor,''
  Class.\ Quant.\ Grav.\  {\bf 25} (2008) 142001
  doi:10.1088/0264-9381/25/14/142001
  [arXiv:0804.2201 [hep-th]]. 
  
  
\bibitem{Bergshoeff:2008ix} 
  E.~A.~Bergshoeff, M.~de Roo, O.~Hohm and D.~Roest,
  ``Multiple Membranes from Gauged Supergravity,''
  JHEP {\bf 0808}, 091 (2008)
  doi:10.1088/1126-6708/2008/08/091
  [arXiv:0806.2584 [hep-th]]. 
  
\bibitem{Bergshoeff:2008bh} 
  E.~A.~Bergshoeff, O.~Hohm, D.~Roest, H.~Samtleben and E.~Sezgin,
  ``The Superconformal Gaugings in Three Dimensions,''
  JHEP {\bf 0809}, 101 (2008)
  doi:10.1088/1126-6708/2008/09/101
  [arXiv:0807.2841 [hep-th]]. 
  
 

  
  
\bibitem{Bagger:2007vi} 
  J.~Bagger and N.~Lambert,
  ``Comments on multiple M2-branes,''
  JHEP {\bf 0802}, 105 (2008)
  doi:10.1088/1126-6708/2008/02/105
  [arXiv:0712.3738 [hep-th]].
  
 
\bibitem{deAzcarraga:2010mr} 
  J.~A.~de Azcarraga and J.~M.~Izquierdo,
  ``n-ary algebras: A Review with applications,''
  J.\ Phys.\ A {\bf 43}, 293001 (2010)
  doi:10.1088/1751-8113/43/29/293001
  [arXiv:1005.1028 [math-ph]]. 
  
\bibitem{Axenides:2008rn} 
  M.~Axenides and E.~Floratos,
  ``Nambu-Lie 3-Algebras on Fuzzy 3-Manifolds,''
  JHEP {\bf 0902}, 039 (2009)
  doi:10.1088/1126-6708/2009/02/039
  [arXiv:0809.3493 [hep-th]].
  
\bibitem{Mukhi:2008ux} 
  S.~Mukhi and C.~Papageorgakis,
  ``M2 to D2,''
  JHEP {\bf 0805}, 085 (2008)
  doi:10.1088/1126-6708/2008/05/085
  [arXiv:0803.3218 [hep-th]].  
  
  \bibitem{ArnoldHydro}
  V.~Arnold, B.~A.~Khesin, ``Topological Methods in Hydrodynamics," 
  Applied Mathematical Sciences 125, Springer. 
  
  


\bibitem{Hohm:2017cey}
  O.~Hohm, V.~Kupriyanov, D.~L\"ust and M.~Traube,
  ``General constructions of L$_{\infty}$ algebras,''
  to be published in Advances in Mathematical Physics, 
  arXiv:1709.10004 [math-ph].
  
\bibitem{Nambu:1973qe} 
  Y.~Nambu,
  ``Generalized Hamiltonian dynamics,''
  Phys.\ Rev.\ D {\bf 7}, 2405 (1973).
  doi:10.1103/PhysRevD.7.2405  
  
\bibitem{Arvanitakis:2015sgs} 
  A.~S.~Arvanitakis,
  ``Higher Spins from Nambu-Chern-Simons Theory,''
  Commun.\ Math.\ Phys.\  {\bf 348}, no. 3, 1017 (2016)
  doi:10.1007/s00220-016-2712-x
  [arXiv:1511.01482 [hep-th]]. 
  
\bibitem{Lee:2014mla}
  K.~Lee, C.~Strickland-Constable and D.~Waldram,
  ``Spheres, generalised parallelisability and consistent truncations,''
  Fortsch.\ Phys.\  {\bf 65} (2017) no.10-11,  1700048
  doi:10.1002/prop.201700048
  [arXiv:1401.3360 [hep-th]].  
  
\bibitem{Baguet:2015sma} 
  A.~Baguet, O.~Hohm and H.~Samtleben,
  ``Consistent Type IIB Reductions to Maximal 5D Supergravity,''
  Phys.\ Rev.\ D {\bf 92}, no. 6, 065004 (2015)
  doi:10.1103/PhysRevD.92.065004
  [arXiv:1506.01385 [hep-th]].  
  
\bibitem{Nicolai:2003bp} 
  H.~Nicolai and H.~Samtleben,
  ``Chern-Simons versus Yang-Mills gaugings in three-dimensions,''
  Nucl.\ Phys.\ B {\bf 668}, 167 (2003)
  doi:10.1016/S0550-3213(03)00569-8
  [hep-th/0303213].  
  
\bibitem{Gomis:2008cv} 
  J.~Gomis, A.~J.~Salim and F.~Passerini,
  ``Matrix Theory of Type IIB Plane Wave from Membranes,''
  JHEP {\bf 0808}, 002 (2008)
  doi:10.1088/1126-6708/2008/08/002
  [arXiv:0804.2186 [hep-th]].
  
\bibitem{Hosomichi:2008qk} 
  K.~Hosomichi, K.~M.~Lee and S.~Lee,
  ``Mass-Deformed Bagger-Lambert Theory and its BPS Objects,''
  Phys.\ Rev.\ D {\bf 78}, 066015 (2008)
  doi:10.1103/PhysRevD.78.066015
  [arXiv:0804.2519 [hep-th]].
  
\bibitem{Bergshoeff:2008ta} 
  E.~A.~Bergshoeff and O.~Hohm,
  ``A Topologically Massive Gauge Theory with 32 Supercharges,''
  Phys.\ Rev.\ D {\bf 78}, 125017 (2008)
  doi:10.1103/PhysRevD.78.125017
  [arXiv:0810.0377 [hep-th]]. 
  
\bibitem{Basu:2004ed} 
  A.~Basu and J.~A.~Harvey,
  ``The M2-M5 brane system and a generalized Nahm's equation,''
  Nucl.\ Phys.\ B {\bf 713}, 136 (2005)
  doi:10.1016/j.nuclphysb.2005.02.007
  [hep-th/0412310].  
  
 \end{thebibliography}
\end{document}